\documentclass[a4paper,fleqn,usenatbib]{mnras}
 \usepackage{lscape}
\usepackage{natbib}
\usepackage[pdftex]{graphicx}
\usepackage{adjustbox}

\DeclareMathVersion{bold}

\def\lsim{\mathrel{\rlap{\lower 3pt\hbox{$\sim$}}\raise 2.0pt\hbox{$<$}}}
\def\gsim{\mathrel{\rlap{\lower 3pt\hbox{$\sim$}} \raise
2.0pt\hbox{$>$}}}
\def\simlt{\mathrel{\rlap{\lower 3pt\hbox{$\sim$}}\raise 2.0pt\hbox{$<$}}}
\def\simgt{\mathrel{\rlap{\lower 3pt\hbox{$\sim$}} \raise 2.0pt\hbox{$>$}}}

\title[Blazars in sub-mm catalogues]{Selecting a complete sample of blazars in sub-millimetre catalogues}
\author[]{
\parbox[t]{\textwidth}
{M. Massardi$^{1,2}$, M. Bonato$^{1,3}$, M. L\'opez-Caniego$^{4}$, V. Galluzzi$^{1,5}$, G. De Zotti$^{3}$, L. Bonavera$^{6,7}$, J. Gonz\'alez-Nuevo$^{6,7}$, A. Lapi$^{2}$, and E. Liuzzo$^{1}$
}\\
\vspace*{8pt} \\	
\parbox[t]{\textwidth}{
$^{1}$INAF - Istituto di Radioastronomia - Italian ALMA Regional Centre, via Gobetti 101, 40129 Bologna, Italy\\
$^{2}$SISSA, Via Bonomea 265, 34136 Trieste, Italy\\ 
$^{3}$INAF-Osservatorio Astronomico di Padova, Vicolo dell'Osservatorio 5, I-35122 Padova, Italy\\
$^{4}$Aurora Technology B.V. for ESA, Camino Bajo del Castillo s/n, Urb. Villafranca del Castillo 28692,
Villanueva de la Canada - Madrid, Spain \\
$^{5}$INAF, Osservatorio Astronomico di Trieste, Via Gian Battista Tiepolo 11, I-34143 Trieste, Italy\\
$^{6}$Departamento de Física, Universidad de Oviedo, C. Federico García Lorca 18, 33007 Oviedo, Spain\\
$^{7}$Instituto Universitario de Ciencias y Tecnologías Espaciales de Asturias (ICTEA), C. Independencia 13, 33004 Oviedo, Spain}
}

\begin{document}

\date{}

%
%\pagerange{\pageref{firstpage}--\pageref{lastpage}} \pubyear{2011}
\pagerange{1--21} \pubyear{}
\maketitle

\begin{abstract}
The \textit{Herschel} Astrophysical Terahertz Large Area Survey (H-ATLAS), that has covered about 642 sq. deg. in 5 bands from 100 to 500 $\mu\rm m$, allows a blind flux-limited selection of blazars at sub-mm wavelengths. However, blazars constitute a tiny fraction of H-ATLAS sources and therefore identifying them is not a trivial task. Using the data on known blazars detected by the H-ATLAS we have defined a locus for 500\,$\mu$m selected blazars and exploited it to select blazar candidates in the H-ATLAS fields. 
Candidates and known blazars in the H-ATLAS equatorial and South Galactic Pole fields were followed up with the Australia Telescope Compact Array (ATCA) or with the Karl G. Jansky Very Large Array (VLA), and matched with existing radio- and mm-catalogues to reconstruct the spectral behaviour over at least 6 orders of magnitude in frequency. We identified a selection approach that, combining the information in the sub-mm and radio domains, efficiently singles out genuine blazars. In this way, we identified a sample of 39 blazars brighter than $S_{500\mu\rm m} = 35\,$mJy in the H-ATLAS fields. Tests made cross-matching the H-ATLAS catalogues with large catalogues of blazar candidates indicate that the sample is complete. The derived counts are compared with model predictions finding good consistency
with the C2Ex model and with estimates based on ALMA data. 
\end{abstract}


\newpage

\section{Introduction} \label{sec:intro}

Blazars are a class of radio loud Active Galactic Nuclei (AGN) whose relativistic jets are pointing within a few degrees of the line-of-sight. Their electromagnetic emission covers an extremely broad frequency range, from radio to $\gamma$-rays. They are playing a crucial role in very high energy ($\gamma$-ray and neutrino) astronomy, a fact that has triggered, in recent years, a great interest in them \citep[e.g.,][]{Massaro2015a, MadejskiSikora2016, Paliya2017, DiMauro2018, FermiLAT2018, IceCubeCollaboration2018, Garrappa2019, Acciari2020}.

The emission spectrum of these sources is coarsely described by two broad peaks, a synchrotron and an inverse-Compton peak \citep[e.g.,][]{Fossati1998, Bottcher2007, Yan2014}. However, very long baseline interferometry (VLBI) images show multiple knots often called ``components'' of the jet \citep[e.g.,][]{Nair2019}. In fact, at a closer analysis, the shape of the synchrotron peak turns out to be quite complex and is explained by the combination of different synchrotron self-absorbed components in a conical geometry \citep[e.g.,][]{Galluzzi2018, Galluzzi2019}. A full understanding of the blazar physics thus demands a multi-frequency study.

The synchrotron peak frequency, in terms of $\nu L_\nu$, varies by several orders of magnitude, from $\sim 10^{12.5}$ to over $10^{18}\,$Hz \citep[e.g.,][]{Giommi2012}. Blazar sub-classes, i.e. flat-spectrum radio quasars (FSRQs) and BL-Lacertae type sources (BL-Lacs), have different distributions of peak frequencies \citep{Fossati1998, Abdo2010, Tucci2011, Giommi2012}, with BL-Lacs reaching much higher values. 

As a consequence, selections at different frequencies emphasize different blazar sub-classes. For example, the radio and the X-ray selections, that have yielded most of known blazars, favour low- and high-frequency peaked sources, respectively. Selections at other frequencies could identify sources peaking at intermediate frequencies, and are thus crucial to achieve a comprehensive view of the blazar population. In this paper we will focus on selection criteria in the sub-millimetre (sub-mm) band.

Source samples blindly selected at mm and sub-mm wavelengths have been provided by \textit{Planck} surveys \citep{Planck2013counts, Planck2018PCNT}. However, picking up the rare blazars among the far more numerous dusty galaxies turned out to be difficult at sub-mm wavelengths, on account of the large (few arcmin) \textit{Planck} positional errors. Moreover, \textit{Planck} photometry has large uncertainties and is liable to contamination by Galactic thermal dust emission or by fluctuations of the Cosmic Infrared Background. Some efforts have been done by the \textit{Planck} Collaboration \citep{2020A&A...644A..99P} to improve on this still open problem. 

The higher sensitivity, higher resolution \textit{Herschel} surveys reach substantially deeper flux densities and largely bypass the contamination problem but still do not allow us to directly single out this rare population. Follow-up observations are necessary to select genuine blazars.

This paper is the third of a series aimed at identifying blazars detected by the \textit{Herschel} Astrophysical Terahertz Large Area Survey \citep[H-ATLAS;][]{Eales2010}, the largest extragalactic survey carried out with the \textit{Herschel} Space Observatory \citep{Pilbratt2010}. The H-ATLAS observations were carried out over 5 high Galactic latitude fields: three equatorial fields around right ascensions of 9\,h, 12\,h and 15\,h, totalling an area of $161.6\,\hbox{deg}^2$ \citep{Valiante2016}, and two fields near the north and the south Galactic poles \citep[NGP and SGP fields;][]{Maddox2018}, with areas of $177.1\,\hbox{deg}^2$ and $303.4\,\hbox{deg}^2$, respectively. The total covered area is thus of $642.1\,\hbox{deg}^2$. The two \textit{Herschel} instruments, the Photodetector Array Camera and Spectrometer \citep[PACS;][]{Poglitsch2010} and the Spectral and Photometric Imaging REceiver \citep[SPIRE;][]{Griffin2010}, operated simultaneously in parallel mode to provide photometry in five bands: 100, 160, 250, 350, and $500\,\mu$m.

\citet{GonzalezNuevo2010} reported the detection of 2 blazars among the 19 candidates selected by cross-matching the H-ATLAS sources with $S_{500\mu\rm m}\ge 50\,$mJy in the $4\,\hbox{deg}\times 4\,\hbox{deg}$ H-ATLAS science demonstration phase field with the Very Large Array (VLA) 1.4 GHz Faint Images of the Radio Sky at Twenty-centimetres (FIRST) catalogue \citep{Becker1995} .

\citet{LopezCaniego2013} extended the search to sources brighter than $S_{500\mu\rm m}= 35\,$mJy over $\simeq 135\,\hbox{deg}^2$ in the H-ATLAS equatorial fields. They cross-matched the H-ATLAS sources in such area with the FIRST catalogue using a 10 arcsec search radius. Then, they looked for counterparts of the 379 sources with FIRST identifications in the 3rd Edition of the Multifrequency Catalogue of Blazars \citep[BZCAT;][]{Massaro2011}, again within 10 arcsec. They found 9 matches.
Other 9 sub-mm blazars were found by cross-matching the 3rd edition of the BZCAT with the \textit{Planck} Early Release Compact Source Catalogue \citep[][containing detections at $545$ and $857\,$GHz (550 and $350\,\mu\rm m$) for these sources]{PlanckERCSC2011}. The distribution in the $\log(S_{500\mu\rm m}/S_{350\mu\rm m})$ versus $\log(S_{500\mu\rm m}/S_{1.4\rm GHz})$ plane of these 18 objects identified a ``blazar region'',
defined by $S_{500\mu\rm m}/S_{1.4\rm GHz}\le 2.4$.

Based on the preliminary H-ATLAS photometry available at the time, \citet{LopezCaniego2013} found twelve sources brighter than $S_{500\mu\rm m}= 35\,$mJy with $S_{500\mu\rm m}/S_{1.4\rm GHz}\le 2.4$, in addition to the 9 BZCAT blazars. The analysis of multi-frequency photometric data from the literature plus new radio observations with the Medicina 32-m telescope at 5 GHz led to the conclusion that only one out of the 12 selected sources has a continuum spectrum compatible with being a blazar.

In this paper we will present the selection of a sample of candidate blazar in all H-ATLAS fields (see Section~ \ref{sect:sample_selection}), and their follow-up (Section~ \ref{sect:radio_obs}) with the Australia Telescope Compact Array (ATCA) in several epochs or with the VLA, and cross-matches with existing catalogues. 

The first observations made clear the reason for the low selection efficiency: the criterion $S_{500\mu{\rm m}}/S_{1.4\rm GHz}\le 2.4$ selects, in addition to blazars with roughly flat spectra from radio to sub-mm wavelengths, also steep-spectrum radio sources hosted by star-forming galaxies whose dust emission dominates at sub-mm wavelengths. 
They also stressed that the spectral energy distributions (SEDs) of blazars strongly differ from those of steep-spectrum dusty sources at frequencies of a few tens of GHz, where the second group show a deep minimum. 

This fact has guided our selection of observing frequencies, and the reconstruction of radio-to-mm spectra. We have acquired ATCA and VLA observations of known and newly selected candidate blazars in various epochs, and cross-matched our samples with the existing radio- and mm- catalogues 
(see sections~\ref{sect:ATCA_obs},~\ref{sect:VLA_obs}, and~\ref{sect:radio_xmatch}), while refining the selection criteria by exploiting the sub-mm colours of sources. 

To this end, we took advantage of the fact that the sample of known blazars detected by the H-ATLAS is now much larger, so that the distribution of their sub-mm colours is much better constrained. The selection could be further honed using the improved, final H-ATLAS photometry and astrometry \citep{Valiante2016,Maddox2018}, and could be extended to the full surveyed area.

This allowed us to identify a better criterion for blazar selection, that can be applied to mm-wave catalogues, and to build a sub-mm selected sample of candidates that can be considered genuine blazars (see Section~\ref{sect:SED}). The SEDs of our sub-mm selected blazars, including the distribution of their synchrotron peak frequencies, are discussed in Sect.~\ref{sect:blazarSEDs}.
We have then analyzed our sample with respect to other selections to quantify, in Section~\ref{sect:comparisons}, the completeness yielded by our criteria. In Sect.~\ref{sec:counts} we work out the blazar number counts at $500\,\mu$m. Our conclusions are summarized in Section~\ref{sec:conclusions}.

Throughout the paper we use the ``combined'' values of cosmological parameters for the base (flat) $\Lambda$CDM model \citep{Planck2018parameters}: $H_0=67.37\,\hbox{km}\,\hbox{s}^{-1}\,\hbox{Mpc}^{-1}$ and $\Omega_m = 0.315$. As for radio spectral indices, $\alpha$, we adopt the convention $S_\nu \propto \nu^\alpha$.

\begin{table*}
\begin{center}
\caption{Sub-mm and radio data for BZCAT blazars in the H-ATLAS equatorial (upper part of the table), SGP (central part) and NGP (bottom part) fields with $S_{500\mu\rm m} > 35\,$mJy. The 100--$500\,\mu$m flux densities and errors are taken from the H-ATLAS catalogues. The 1.4\,GHz flux densities, $S_{1.4}$, are from the FIRST catalogue in the case of equatorial and NGP fields, and from the NVSS catalogue for the SGP field. The source IAU names and coordinates (RA and Dec in degrees) are from the
H-ATLAS catalogues. All flux densities, denoted by ``S'', and their errors, denoted by ``E'' are 
in mJy/beam. In the FIRST case, $S_{1.4}$ is the peak flux density. Spectroscopic redshifts $z_{\rm H}$ and $z_{\rm B}$
are from the H-ATLAS or literature and from the BZCAT catalogues, respectively; ``$...$'' means that a value is missing. The angular separations (``sep'') between the \textit{Herschel} and the radio positions are in arcsec.
The indexes $A$, $B$, $C$, and $V$ here and in the following table indicate sources observed during dedicated ATCA, as presented in Table~\ref{tab:ATCAEpochs}, and VLA follow-up observations.
}\label{tab:BZCAT}
\hspace{-0.8cm}
\resizebox{1.04\textwidth}{!}{
\setlength{\tabcolsep}{2pt}
\begin{tabular}{|r|l|r|r|r|r|r|r|r|r|r|r|r|r|l|l|r|r|r|}
\hline
  \multicolumn{1}{|c|}{No.} &
  \multicolumn{1}{|c|}{IAU ID} &
  \multicolumn{1}{c|}{RA} &
  \multicolumn{1}{c|}{Dec} &
  \multicolumn{1}{c|}{S$_{100}$} &
  \multicolumn{1}{c|}{E$_{100}$} &
  \multicolumn{1}{c|}{S$_{160}$} &
  \multicolumn{1}{c|}{E$_{160}$} &
  \multicolumn{1}{c|}{S$_{250}$} &
  \multicolumn{1}{c|}{E$_{250}$} &
  \multicolumn{1}{c|}{S$_{350}$} &
  \multicolumn{1}{c|}{E$_{350}$} &
  \multicolumn{1}{c|}{S$_{500}$} &
  \multicolumn{1}{c|}{E$_{500}$} &
  \multicolumn{1}{c|}{$z_{\rm H}$} &
  \multicolumn{1}{c|}{$z_{\rm B}$} &
  \multicolumn{1}{c|}{S$_{1.4}$} &
  \multicolumn{1}{c|}{E$_{1.4}$} &
  \multicolumn{1}{c|}{sep} \\
\hline
1	&	J083949.3+010437 	&	 129.95574 	&	 +1.07706 	&	   31.8 	&	 45.4 	&	  74.3 	&	 50.3 	&	  49.4 	&	 7.5 	&	  44.2 	&	 8.3 	&	  42.6 	&	 8.7 	&	 0.33745 	&	1.123	&	 443.7 	&	 0.15 	&	 11.09\\
2	&	J090910.1+012134$^B$ 	&	 137.29236 	&	 +1.35968 	&	  180.0 	&	 41.3 	&	  98.9 	&	 47.9 	&	 256.5 	&	 6.4 	&	 327.0 	&	 7.4 	&	 375.3 	&	 8.1 	&	 1.0245 	&	1.024	&	 559.6 	&	 0.14 	&	 1.49\\
3	&	J090940.2+020004$^B$ 	&	 137.41775 	&	 +2.00113 	&	 24.5 	&	 45.4 	&	 0.0 	&	 50.3 	&	30.4 	&	 6.6 	&	41.8 	&	 7.5 	&	62.3 	&	 7.8 	&	 1.7468 	&	 $...$ 	&	 305.8 	&	 0.15 	&	 6.27\\
4	&	J113245.7+003427$^B$ 	&	 173.19058 	&	 +0.57429 	&	 29.5 	&	 41.3 	&	  24.1 	&	 47.9 	&	66.8 	&	 7.4 	&	72.3 	&	 8.4 	&	63.1 	&	 8.9 	&	 1.2227 	&	1.223	&	 469.0 	&	 0.15 	&	 1.58\\
5	&	J113302.8+001545$^B$ 	&	 173.26199 	&	 +0.26263 	&	 13.2 	&	 41.3 	&	  67.0 	&	 47.9 	&	52.0 	&	 7.2 	&	45.3 	&	 8.2 	&	38.3 	&	 8.6 	&	 1.1703 	&	1.173	&	 214.9 	&	 0.15 	&	 4.28\\
6	&	J113320.2+004053$^B$ 	&	 173.33418 	&	 +0.68160 	&	 24.7 	&	 41.3 	&	  32.1 	&	 47.9 	&	37.2 	&	 7.5 	&	43.0 	&	 8.2 	&	43.8 	&	 8.7 	&	 $...$ 	&	1.633	&	 312.5 	&	 0.16 	&	 2.15\\
7	&	J115043.6$-$002354$^B$ 	&	 177.68174 	&	 $-$0.39845 	&	  101.5 	&	 41.3 	&	  84.3 	&	 47.9 	&	36.5 	&	 7.4 	&	56.7 	&	 8.0 	&	83.0 	&	 8.5 	&	 1.9796 	&	1.976	&	  2803.2 	&	 0.63 	&	 4.17\\
8	&	J120741.7$-$010636$^B$ 	&	 181.92379 	&	 $-$1.11008 	&	7.8 	&	 45.4 	&	0.0 	&	 50.3 	&	25.0 	&	 7.0 	&	28.1 	&	 8.3 	&	44.1 	&	 8.6 	&	 $...$ 	&	1.006	&	79.1 	&	 0.14 	&	 1.11\\
9	&	J121758.7$-$002946$^B$ 	&	 184.49498 	&	 $-$0.49635 	&	 72.2 	&	 41.3 	&	  62.2 	&	 47.9 	&	 122.7 	&	 7.4 	&	 152.3 	&	 8.1 	&	 177.5 	&	 8.5 	&	 0.4190 	&	0.419	&	 239.2 	&	 0.14 	&	 2.11\\
10	&	J121834.9$-$011954$^B$ 	&	 184.64555 	&	 $-$1.33174 	&	2.1 	&	 41.3 	&	  24.8 	&	 47.9 	&	35.8 	&	 7.3 	&	59.2 	&	 8.1 	&	52.1 	&	 8.9 	&	 0.1683 	&	 $...$ 	&	 274.1 	&	 0.15 	&	 0.58\\
11	&	J141004.7+020306$^B$ 	&	 212.51964 	&	 +2.05187 	&	 54.5 	&	 41.3 	&	  75.6 	&	 47.9 	&	 119.4 	&	 7.3 	&	 151.0 	&	 8.3 	&	 176.0 	&	 8.7 	&	 0.1801 	&	 $...$ 	&	 291.6 	&	 0.15 	&	 1.17\\
\hline																																				
12	&	J001035.7$-$302745$^B$ 	&	2.64865 	&	 $-$30.46251 	&	 0.0 	&	 38.7 	&	8.4 	&	 41.7 	&	28.8 	&	 7.5 	&	35.4 	&	 8.4 	&	36.1 	&	 9.1 	&	 $...$ 	&	1.190	&	 315.3 	&	9.5 	&	 0.86\\
13	&	J001941.9$-$303116 	&	 4.92452 	&	 $-$30.52111 	&	 74.5 	&	 41.8 	&	 18.4 	&	 44.1 	&	 20.8 	&	 6.8 	&	 17.6 	&	 8.4 	&	 36.1 	&	 8.7 	&	 $...$ 	&	2.677	&	 507.0	&	 15.2 	&	 10.64 \\
14	&	J003233.0$-$284921 	&	8.13748 	&	 $-$28.82237 	&	 42.7 	&	 41.8 	&	 0.0 	&	 44.0 	&	59.2 	&	 7.7 	&	60.5 	&	 8.4 	&	54.0 	&	 8.5 	&	 $...$ 	&	0.324	&	 159.9 	&	5.7 	&	 2.06\\
15	&	J005802.3$-$323420$^A$ 	&	 14.50938 	&	 $-$32.57215 	&	 76.8 	&	 37.1 	&	 65.3 	&	 40.5 	&	60.6 	&	 6.9 	&	61.5 	&	 7.8 	&	70.0 	&	 8.1 	&	 $...$ 	&	$...$	&	 185.9 	&	5.6 	&	 1.46\\
16	&	J014310.0$-$320056$^A$ 	&	 25.79166 	&	 $-$32.01567 	&	 0.0 	&	 41.8 	&	  102.5 	&	 44.0 	&	96.0 	&	 7.5 	&	 119.5 	&	 8.4 	&	 122.4 	&	 9.0 	&	 $...$ 	&	0.375	&	75.8 	&	2.3 	&	 2.08\\
17	&	J014503.4$-$273333 	&	 26.26398 	&	 $-$27.55915 	&	0.0 	&	 41.8 	&	0.0 	&	 44.0 	&	 131.4 	&	 7.8 	&	 179.2 	&	 8.8 	&	 234.4 	&	 9.0 	&	 $...$ 	&	1.148	&	 922.8 	&	 27.7 	&	 2.00\\
18	&	J224838.6$-$323551$^B$ 	&	  342.16084 	&	 $-$32.59745 	&	 77.4 	&	 36.1 	&	 91.1 	&	 39.8 	&	 119.2 	&	 7.7 	&	 152.8 	&	 8.3 	&	 194.7 	&	 8.6 	&	 $...$ 	&	2.268	&	 707.6 	&	 21.2 	&	 0.95\\
19	&	J231448.5$-$313837$^A$ 	&	  348.70220 	&	 $-$31.64348 	&	0.0 	&	 38.7 	&	 98.3 	&	 41.7 	&	53.5 	&	 7.7 	&	55.3 	&	 8.3 	&	50.8 	&	 8.6 	&	 $...$ 	&	1.323	&	 825.2 	&	 24.8 	&	 2.37\\
20	&	J235347.4$-$303746$^B$ 	&	  358.44757 	&	 $-$30.62938 	&	 31.7 	&	 41.8 	&	 49.0 	&	 44.0 	&	77.1 	&	 7.4 	&	96.6 	&	 8.4 	&	 103.1 	&	 8.9 	&	 $...$ 	&	0.737	&	 397.0 	&	 14.0 	&	 2.53\\
21	&	J235935.3$-$313343$^B$ 	&	  359.89723 	&	 $-$31.56206 	&	 70.9 	&	 41.8 	&	 83.2 	&	 44.0 	&	61.2 	&	 7.7 	&	67.8 	&	 8.4 	&	83.8 	&	 9.2 	&	 $...$ 	&	0.990	&	 347.4 	&	 10.4 	&	 2.30\\
\hline																																				
22	&	J125757.3+322930 	&	 194.48882 	&	 +32.49176 	&	  306.2 	&	 39.1 	&	  337.8 	&	 42.0 	&	 143.7 	&	 7.4 	&	 188.4 	&	 8.3 	&	 214.9 	&	 8.6 	&	 0.8052 	&	0.805	&	 589.2 	&	  0.14 	&	 1.45\\
23	&	J130129.0+333703 	&	 195.37091 	&	 +33.61742 	&	0.0 	&	 43.6 	&	 0.0 	&	 45.4 	&	31.5 	&	 7.4 	&	38.9 	&	 8.2 	&	40.6 	&	 8.5 	&	 1.0084 	&	1.009	&	71.2 	&	  0.15 	&	 2.82\\
24	&	J131028.7+322044 	&	 197.61967 	&	 +32.34551 	&	  702.6 	&	 43.6 	&	  681.3 	&	 45.4 	&	259.1 	&	 6.8 	&	 363.1 	&	 7.8 	&	 452.3 	&	 8.1 	&	 0.9959 	&	0.997	&	  1459.0 	&	  0.32 	&	 0.66\\
25	&	J131059.2+323331 	&	 197.74674 	&	 +32.55867 	&	 0.0 	&	 43.6 	&	 60.1 	&	 45.4 	&	37.6 	&	 7.5 	&	63.8 	&	 8.3 	&	82.3 	&	 8.4 	&	 1.6391 	&	1.639	&	 243.4 	&	  0.30 	&	 4.11\\
26	&	J131443.7+234828 	&	 198.68208 	&	 +23.80790 	&	 99.2 	&	 43.6 	&	3.7 	&	 45.4 	&	49.6 	&	 7.4 	&	62.6 	&	 8.4 	&	61.5 	&	 8.6 	&	 0.2256 	&	$...$	&	 169.0 	&	  0.13 	&	 2.20\\
27	&	J131736.4+342518 	&	 199.40172 	&	 +34.42168 	&	 83.0 	&	 43.6 	&	 79.3 	&	 45.4 	&	77.1 	&	 7.3 	&	99.5 	&	 8.1 	&	 112.0 	&	 8.6 	&	 1.0542 	&	1.055	&	 421.0 	&	  0.15 	&	 2.45\\
28	&	J132248.0+321607 	&	 200.69983 	&	 +32.26859 	&	 3.7 	&	 43.6 	&	 53.0 	&	 45.4 	&	 31.1 	&	 7.0 	&	 33.1 	&	 7.8 	&	 35.0 	&	 8.0 	&	 1.3876 	&	 $...$ 	&	 394.6 	&	 0.15 	&	 7.29\\
29	&	J132952.9+315410 	&	 202.47026 	&	 +31.90275 	&	 18.0 	&	 43.6 	&	 72.7 	&	 45.4 	&	50.4 	&	 7.4 	&	71.0 	&	 8.1 	&	86.4 	&	 9.0 	&	$...$ 	&	$...$	&	 779.0 	&	  0.46 	&	 1.34\\
30	&	J133307.4+272518 	&	 203.28080 	&	 +27.42166 	&	 0.0 	&	 43.6 	&	 0.0 	&	 45.4 	&	89.3 	&	 7.5 	&	 104.6 	&	 8.2 	&	 117.1 	&	 8.3 	&	 0.7275 	&	2.126	&	 209.5 	&	  0.14 	&	 1.29\\
31	&	J134208.4+270933 	&	 205.53514 	&	 +27.15913 	&	 0.0 	&	 43.6 	&	 90.1 	&	 45.4 	&	48.9 	&	 7.5 	&	48.9 	&	 8.3 	&	50.1 	&	 8.5 	&	 1.1895 	&	1.185	&	 238.1 	&	  0.14 	&	 2.29\\
\hline
\end{tabular}       }
\end{center}
\end{table*}

\begin{figure*}
\begin{center}
\includegraphics[width=0.8\textwidth]{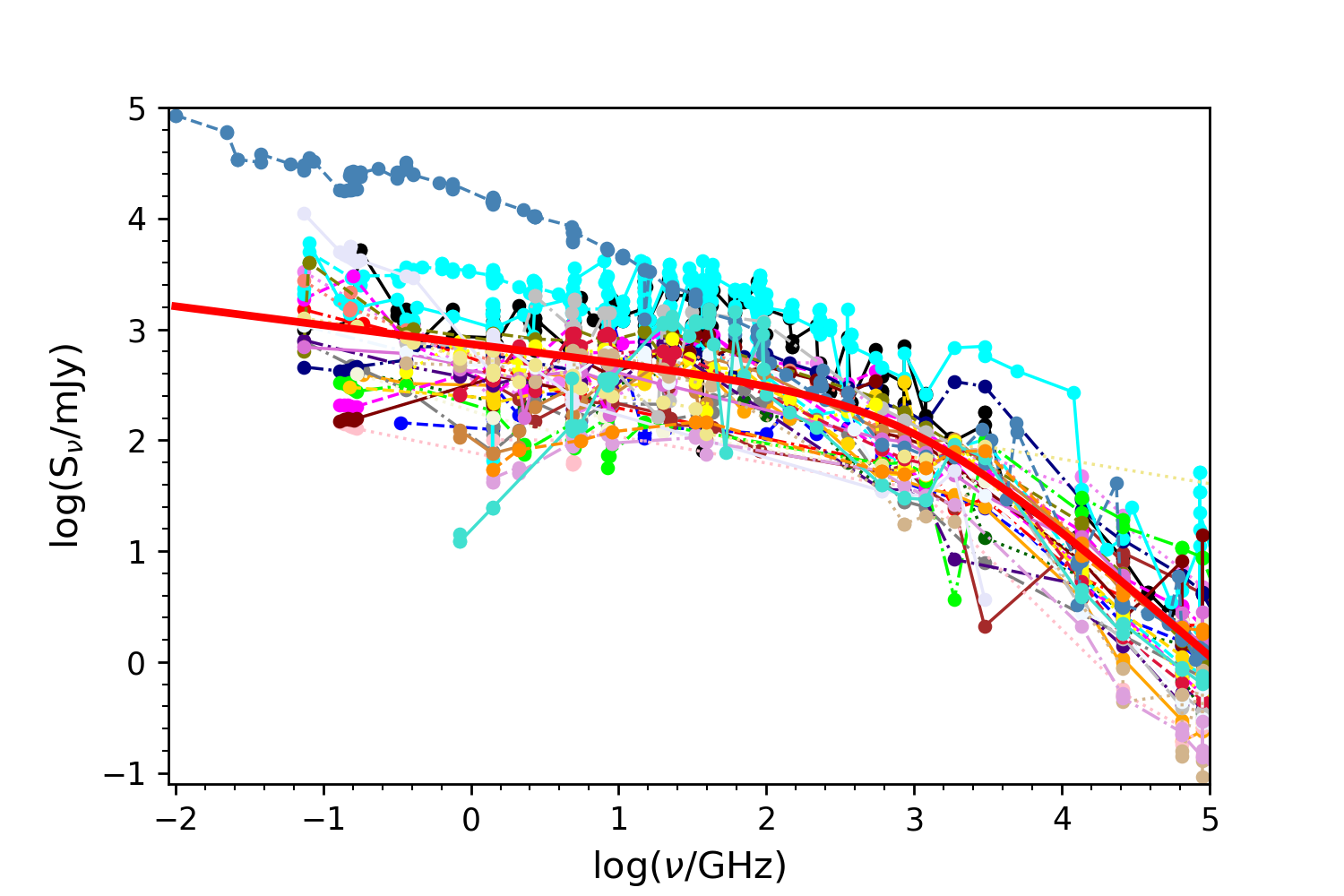}
\end{center}
\caption{Observed SEDs of the BZCAT blazars listed in Table \ref{tab:BZCAT} with
$S_{500\mu\rm m}> 35\,$mJy in the H-ATLAS fields. The thick solid red line shows the smoothed average SED of these blazars (see text). SEDs of individual sources, in terms of $\nu L_\nu$, are available as Supplementary Material. One example is shown in Fig.~\ref{fig:example}. }\label{fig:blazarSEDs}
%The data sources are the same as in Fig.~\ref{fig:SEDs} except that the brightest objects also have photometry from the \textit{Planck} multi-frequency catalogue of non-thermal sources \citep[PCNT;][]{Planck2018PCNT}, extracted using a search radius of $3'$. Six blazars in equatorial fields have GMRT flux densities at 325\,MHz \citep{Mauch2013}; we have taken the peak values. 
\end{figure*}

\begin{figure}
\begin{center}
\includegraphics[width=0.48\textwidth]{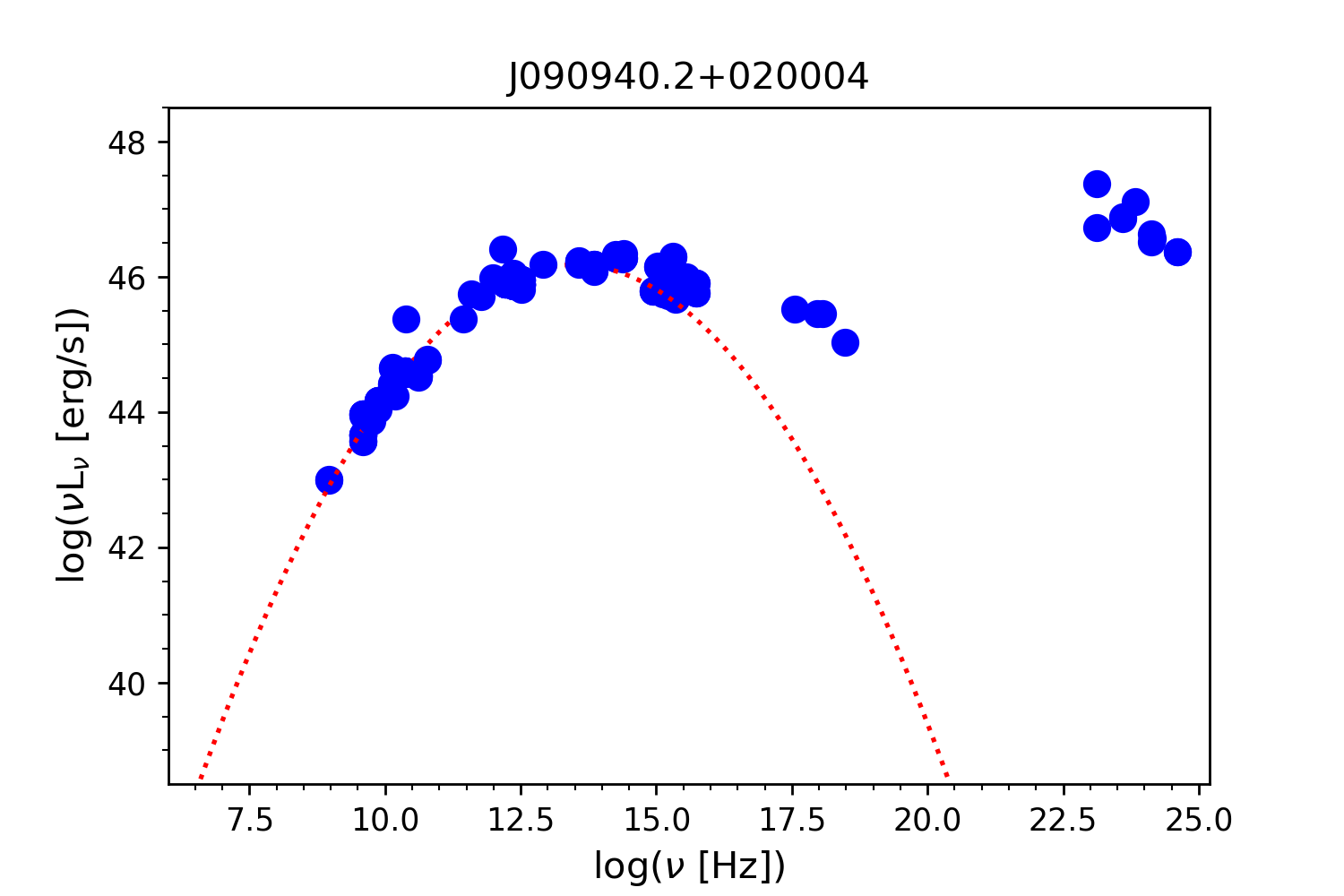}
\end{center}
\caption{SED, in terms of $\nu L_\nu$, of the blazar J090940.2+020004. Data points have been extracted from the ASI Space Science Data Center (ASI-SSDC) and the NASA Extragalactic Database (NED) complemented with the H-ATLAS data and our radio observations.}
\label{fig:example}
\end{figure}

\begin{table*}
\begin{center}
\caption{\textit{Herschel} and FIRST or NVSS (for SGP sources) data on the equatorial (upper part of the table), SGP (second part) and NGP (third part) H-ATLAS blazar candidates selected as specified in the text, excluding the known BZCAT blazars listed in Table~\ref{tab:BZCAT}. The source IAU names and coordinates (RA and Dec in degrees) are from the H-ATLAS catalogues. The label in the last column indicates if our follow-up recognized the source as a blazar (``bl''), or not (``cand'').
In the bottom part of the table we listed the 16 sources selected on the basis of earlier selection criteria and/or earlier versions of the H-ATLAS catalogue, which don't obey the latest criteria applied to the final version of the H-ATLAS catalogue. These 16 sources constitute a control sample to check whether our final criteria are too restrictive. We flagged them with an asterisk in the classification column in this and the following tables.
The other symbols have the same meaning as in Table~\protect\ref{tab:BZCAT}.}\label{tab:candidates}
\setlength{\tabcolsep}{2pt}
\begin{tabular}{llcccccccccccccccc} 
\hline
No. & IAU ID &RA &  Dec & 
   S$_{100}$ &
   E$_{100}$ &
   S$_{160}$ &
   E$_{160}$ &
   S$_{250}$ &
   E$_{250}$ &
   S$_{350}$ &
   E$_{350}$ &
   S$_{500}$ &
   E$_{500}$ &
   S$_{1.4}$ &
   E$_{1.4}$ &
   sep &
   class
   \\
\hline
1	&	J120123.1+002830$^B$ 	&	 180.34660 	&	 +0.47506 	&	 46.5 	&	 41.3 	&	 3.4 	&	 47.9 	&	34.7 	&	 7.3 	&	38.7 	&	 8.2 	&	37.6 	&	 8.6 	&	 311.3 	&	 0.23 	&	 1.21	&	bl\\
2	&	 J141015.9+005430$^B$ 	&	 212.56635 	&	 +0.90849  	&	   1.5 	&	 45.4 	&	   ... 	&	 50.3 	&	  25.7 	&	 7.2 	&	  39.9 	&	 8.3 	&	  43.4 	&	 8.9 	&	  41.4 	&	 0.15 	&	 2.18	&	 cand\\
3	&	 J145146.1+010608$^A$ 	&	 222.94249 	&	 +1.10249  	&	 ... 	&	 45.4 	&	 26.3 	&	 50.3 	&	 31.5 	&	 7.2  	&	 38.6 	&	 8.1 	&	 52.7 	&	 8.4 	&	 41.8 	&	 0.15 	&	 10.78 	&	 bl \\
\hline																																		
4	&	 J000438.2$-$321059$^{A,C}$ 	&	   1.15929 	&	 $-$32.18314 	&	  82.2 	&	 41.8 	&	 178.2 	&	 44.0 	&	  76.6 	&	 7.5 	&	   75.3 	&	 8.4 	&	  67.6 	&	 9.0 	&	  69.7 	&	 2.1 	&	 2.46	&	 cand\\
5	&	 J000935.6$-$321639$^{A}$ 	&	   2.39836 	&	 $-$32.27763 	&	  88.2 	&	 38.7 	&	  86.5 	&	 41.7 	&	  68.1 	&	 7.5 	&	   72.5 	&	 8.3 	&	  60.0 	&	 8.7 	&	 388.3 	&	 11.7 	&	 3.22 	&	 bl \\
6	&	 J002213.1$-$302034$^C$ 	&	   5.55475 	&	 $-$30.34278 	&	  34.9 	&	 41.8 	&	   ... 	&	 44.0 	&	  29.2 	&	 7.4 	&	   31.7 	&	 8.3 	&	  39.8 	&	 8.5 	&	  44.3 	&	 1.8 	&	 6.37	&	 cand\\
7	&	 J002615.8$-$351247 	&	   6.56571 	&	 $-$35.21295 	&	  69.7 	&	 41.8 	&	  90.9 	&	 44.0 	&	  29.1 	&	 7.4 	&	   30.3 	&	 8.5 	&	  39.6 	&	 8.7 	&	  24.6 	&	 0.9 	&	 7.22 	&	 bl\\
8	&	 J003246.2$-$293109$^C$ 	&	   8.19239 	&	 $-$29.51917 	&	   ... 	&	 41.8 	&	  40.2 	&	 44.0 	&	  57.5 	&	 7.2 	&	   58.6 	&	 8.2 	&	  56.9 	&	 8.6 	&	 572.0 	&	 17.2 	&	 0.79	&	 cand\\
9	&	 J010726.5$-$310822$^{B,C}$ 	&	  16.86023 	&	 $-$31.13957 	&	   ... 	&	 41.8 	&	  19.8 	&	 44.0 	&	   29.0 	&	 7.7 	&	   41.6 	&	 8.3 	&	  41.2 	&	 8.8 	&	  34.2 	&	 1.1 	&	 6.68	&	 cand\\
10	&	 J011721.2$-$310845$^C$ 	&	  19.33836 	&	 $-$31.14587 	&	  79.9 	&	 41.8 	&	  79.5 	&	 44.0 	&	   56.7 	&	 7.5 	&	   49.7 	&	 8.2 	&	  52.5 	&	 8.9 	&	  54.0 	&	 2.1 	&	 4.08 	&	 bl \\
11	&	 J013029.6$-$324847$^{A,C}$ 	&	  22.62342 	&	 $-$32.81308 	&	  12.0 	&	 41.8 	&	   ... 	&	 44.0 	&	   52.2 	&	 6.8 	&	   53.5 	&	 7.5 	&	  44.4 	&	 7.9 	&	  29.7 	&	 1.3 	&	 4.80	&	 cand\\
12	&	 J222321.6$-$313701$^{A,C}$ 	&	 335.83998 	&	 $-$31.61687 	&	  32.3 	&	 38.7 	&	  36.3 	&	 41.7 	&	   86.0 	&	 9.5 	&	  110.9 	&	10.5 	&	 131.9 	&	11.7 	&	 220.3 	&	 6.6 	&	 1.06 	&	 bl \\
13	&	 J223044.9$-$290038$^C$ 	&	 337.68721 	&	 $-$29.01060 	&	   ... 	&	 41.8 	&	  45.3 	&	 44.0 	&	   44.2 	&	 9.4 	&	   52.6 	&	10.0 	&	  55.3 	&	10.1 	&	  57.4 	&	 1.8 	&	 7.63	&	 cand\\
14	&	 J223454.5$-$325910$^{A,C}$ 	&	 338.72725 	&	 $-$32.98624 	&	   ... 	&	 38.7 	&	  16.9 	&	 41.7 	&	   35.3 	&	 7.4 	&	   41.9 	&	 8.4 	&	  40.8 	&	 9.3 	&	  66.0 	&	 2.0 	&	 2.54	&	 cand\\
15	&	 J224041.0$-$335946$^{B,C}$ 	&	 340.17100 	&	 $-$33.99624 	&	  24.1 	&	 37.1 	&	   ... 	&	 40.5 	&	   45.5 	&	 7.5 	&	   57.0 	&	 8.4 	&	  48.1 	&	 9.2 	&	  64.3 	&	 2.0 	&	 8.66	&	 cand\\
16	&	 J225036.6$-$323909$^C$ 	&	 342.65243 	&	 $-$32.65245 	&	  34.7 	&	 36.1 	&	  35.5 	&	 39.8 	&	   35.2 	&	 7.3 	&	   37.4 	&	 8.3 	&	  36.0 	&	 8.4 	&	  23.4 	&	 0.8 	&	  9.44	&	 cand\\
17	&	 J231859.8$-$294738$^{B,C}$ 	&	 349.74920 	&	 $-$29.79385 	&	  24.4 	&	 38.7 	&	  14.7 	&	 41.7 	&	   44.7 	&	 9.1 	&	   53.8 	&	 9.4 	&	  57.6 	&	10.2 	&	  93.9 	&	 2.9 	&	 3.41	&	 cand\\
18	&	 J232643.9$-$350420$^{B,C}$ 	&	 351.68309 	&	 $-$35.07214 	&	  73.6 	&	 41.8 	&	   ... 	&	 44.0 	&	   35.2 	&	 6.9 	&	   46.1 	&	 7.9 	&	  41.2 	&	 8.2 	&	  33.6 	&	 1.1 	&	 6.04	&	 cand\\
19	&	 J234641.2$-$304858$^{B,C}$ 	&	 356.67169 	&	 $-$30.81617 	&	   ... 	&	 41.8 	&	   ... 	&	 44.0 	&	   26.6 	&	 7.4 	&	   47.3 	&	 8.4 	&	  39.6 	&	 8.8 	&	  42.5 	&	 1.4 	&	 7.34	&	 cand\\
20	&	 J234934.3$-$295254$^C$ 	&	 357.39311 	&	 $-$29.88161 	&	   ... 	&	 41.8 	&	  98.4 	&	 44.0 	&	   46.6 	&	 7.5 	&	   42.1 	&	 8.2 	&	  35.3 	&	 8.5 	&	  71.7 	&	 2.2 	&	 8.70	&	 cand\\
\hline																																		
21	&	 J124735.9+292644 	&	 191.89947 	&	 +29.44549 	&	   ... 	&	 43.6 	&	  65.9 	&	 45.4 	&	  48.1 	&	 7.7 	&	  48.1 	&	 8.3 	&	  41.3 	&	 8.6 	&	   151.2 	&	 0.13 	&	 3.63	&	 cand\\
22	&	 J125848.3+231602 	&	 194.70145 	&	 +23.26719 	&	  32.3 	&	 43.6 	&	  70.2 	&	 45.4 	&	  34.1 	&	 7.6 	&	  30.2 	&	 8.3 	&	  42.1 	&	 8.5 	&	    33.9 	&	 0.14 	&	 7.38	&	 cand\\
23	&	 J131213.2+244754 	&	 198.05485 	&	 +24.79823 	&	   ... 	&	 43.6 	&	  14.7 	&	 45.4 	&	  44.1 	&	 7.3 	&	  60.4 	&	 8.2 	&	  50.9 	&	 8.3 	&	    56.8 	&	 0.14 	&	 2.99	&	 cand\\
24	&	 J132301.1+294146 	&	 200.75438 	&	 +29.69622 	&	  95.2 	&	 43.6 	&	  26.9 	&	 45.4 	&	  30.9 	&	 7.4 	&	  32.7 	&	 8.3 	&	  35.0 	&	 8.4 	&	   394.6 	&	 0.16 	&	 2.58	&	 cand\\
25	&	J133038.1+250901 	&	 202.65856 	&	 +25.15023 	&	 116.5 	&	 43.6 	&	 18.5 	&	 45.4 	&	 34.9 	&	 7.4 	&	 41.4 	&	 8.4 	&	 49.3 	&	 8.9 	&	  6826.4 	&	 1.2 	&	 11.16	&	bl\\
26	&	J133108.4+303034 	&	 202.78517 	&	 +30.50945 	&	 61.6 	&	 43.6 	&	 70.3 	&	 45.4 	&	71.8 	&	 7.6 	&	87.1 	&	 8.2 	&	92.2 	&	 8.5 	&	  14774.4 	&	 1.77 	&	 2.28	&	bl\\
27	&	 J133245.8+291047 	&	 203.19093 	&	 +29.17985 	&	  11.9 	&	 40.7 	&	  16.9 	&	 43.1 	&	  43.2 	&	 6.9 	&	  43.5 	&	 7.7 	&	  40.7 	&	 8.2 	&	   108.6 	&	 0.15 	&	 6.58	&	 cand\\
28	&	 J133940.4+303312 	&	 204.91837 	&	 +30.55342 	&	  14.0 	&	 43.6 	&	  31.1 	&	 45.4 	&	  33.9 	&	 7.4 	&	  43.2 	&	 8.3 	&	  35.8 	&	 8.7 	&	    35.4 	&	 0.14 	&	 5.75	&	 cand\\
\hline
29	&	 J085419.0$-$003613$^{B,V}$ 	&	 133.57944 	&	 $-$0.60375 	&	  42.3 	&	 45.4 	&	  45.4 	&	 50.3 	&	  49.0 	&	 7.4 	&	  48.4 	&	 8.1 	&	  37.3 	&	 8.8 	&	  25.4 	&	 0.15 	&	 7.94	&	 cand*\\
30	&	 J120323.5$-$015846$^{B,V}$ 	&	 180.84806 	&	 $-$1.97965 	&	   5.7 	&	 45.4 	&	  75.4 	&	 50.3 	&	  34.8 	&	 7.4 	&	  50.1 	&	 8.3 	&	  39.0 	&	 8.6 	&	 107.4 	&	 0.15 	&	 8.46	&	 cand*\\
31	&	 J121008.3+011528$^{B,V}$ 	&	 182.53482 	&	 +1.25803  	&	  43.3 	&	 45.4 	&	  39.8 	&	 50.3 	&	  59.8 	&	 7.3 	&	  59.9 	&	 8.1 	&	  39.3 	&	 8.8 	&	 136.0 	&	 0.15 	&	 3.48	&	 cand*\\
32	&	 J142631.1+014228$^A$ 	&	 216.62972 	&	 +1.70779  	&	   ... 	&	 32.1 	&	  38.5 	&	 35.6 	&	  30.5 	&	 7.2 	&	  37.6 	&	 8.1 	&	  40.3 	&	 8.5 	&	  23.4 	&	 0.14 	&	 5.20	&	 cand*\\
33	&	 J000348.6$-$343043$^{A,C}$ 	&	   0.95251 	&	 $-$34.51191 	&	   ... 	&	 41.8 	&	  99.6 	&	 44.0 	&	  71.6 	&	 7.6 	&	   63.2 	&	 8.3 	&	  45.8 	&	 8.5 	&	  68.6 	&	 2.1 	&	 2.72	&	 cand*\\
34	&	 J001551.6$-$322927$^{A,C}$ 	&	   3.96492 	&	 $-$32.49093 	&	  80.3 	&	 41.8 	&	  62.7 	&	 44.0 	&	  65.1 	&	 6.7 	&	   62.4 	&	 7.5 	&	  50.9 	&	 7.8 	&	  34.0 	&	 1.1 	&	 1.61	&	 cand*\\
35	&	 J002619.6$-$355344$^C$ 	&	   6.58187 	&	 $-$35.89548 	&	  25.0 	&	 41.8 	&	   9.3 	&	 44.0 	&	  30.4 	&	 9.2 	&	   12.1 	&	10.1 	&	  47.2 	&	10.7 	&	  77.6 	&	 2.4 	&	 9.55 	&	 cand*\\
36	&	 J003959.4$-$303333$^{A,C}$ 	&	   9.99766 	&	 $-$30.55923 	&	   ... 	&	 41.8 	&	  71.8 	&	 44.0 	&	  53.8 	&	 6.9 	&	   61.8 	&	 7.8 	&	  47.6 	&	 8.2 	&	  58.6 	&	 1.8 	&	 9.70	&	 cand*\\
37	&	 J224500.0$-$343029$^{B,C}$ 	&	 341.25009 	&	 $-$34.50818 	&	  31.5 	&	 37.1 	&	 102.5 	&	 40.5 	&	   85.0 	&	 7.5 	&	   93.9 	&	 8.5 	&	  68.9 	&	 9.2 	&	 396.6 	&	 11.9 	&	 2.40	&	 cand*\\
38	&	 J232125.5$-$315637$^{B,C}$ 	&	 350.35623 	&	 $-$31.94352 	&	   ... 	&	 38.7 	&	  33.2 	&	 41.7 	&	   53.5 	&	 7.7 	&	   60.6 	&	 8.2 	&	  41.6 	&	 8.8 	&	 22.8 	&	 0.8 	&	 2.84	&	 cand*\\
39	&	 J235009.9$-$293948$^C$ 	&	 357.54116 	&	 $-$29.66335 	&	  ... 	&	 ... 	&	  ... 	&	 ... 	&	   90.1 	&	 7.3 	&	  115.6 	&	 8.5 	&	  73.4 	&	 8.8 	&	  70.6 	&	 2.2 	&	 9.83	&	 cand*\\
40	&	 J125046.5+263848 	&	 192.69365 	&	 +26.64653 	&	  55.2 	&	 43.6 	&	  55.1 	&	 45.4 	&	  53.3 	&	 7.5 	&	  57.6 	&	 8.2 	&	  42.2 	&	 8.7 	&	    42.6 	&	 0.14 	&	 5.76	&	 cand*\\
41	&	 J125138.0+241126 	&	 192.90826 	&	 +24.19063 	&	   ... 	&	 43.6 	&	  23.8 	&	 45.4 	&	  49.0 	&	 7.4 	&	  50.9 	&	 8.2 	&	  39.2 	&	 8.9 	&	    19.6 	&	 0.14 	&	 1.05	&	 cand*\\
42	&	 J125931.0+250449 	&	 194.87907 	&	 +25.08039 	&	   ... 	&	 43.6 	&	   2.2 	&	 45.4 	&	  81.4 	&	 7.5 	&	  74.4 	&	 8.2 	&	  50.5 	&	 8.7 	&	    37.1 	&	 0.14 	&	 1.13	&	 cand*\\
43	&	 J130724.1+290227 	&	 196.85030 	&	 +29.04075 	&	   ... 	&	 39.1 	&	  53.8 	&	 42.0 	&	  39.5 	&	 6.9 	&	  46.5 	&	 7.7 	&	  36.8 	&	 8.2 	&	    18.8 	&	 0.10 	&	 2.91	&	 cand*\\
44	&	 J133924.4+303257 	&	 204.85168 	&	 +30.54923 	&	  84.0 	&	 43.6 	&	  23.8 	&	 45.4 	&	  83.5 	&	 7.5 	&	  84.3 	&	 8.2 	&	  66.3 	&	 8.5 	&	    75.5 	&	 0.15 	&	 4.93	&	 cand*\\
\hline
\end{tabular}           
\end{center}
\end{table*}
\section{Sample selection}\label{sect:sample_selection}

To define improved selection criteria we first cross-matched the H-ATLAS catalogues\footnote{\url{https://www.h-atlas.org/public-data/download}} with the latest (5th) edition of the BZCAT \citep{Massaro2015b}, considering only sources with $S_{500\mu\rm m}> 35\,$mJy and using a search radius of $10\,$arcsec. We found 29 matches (10 in the equatorial fields, 10 in the NGP field, 9 in the SGP field).

The H-ATLAS source detection and the determination of source positions were made using the $250\,\mu$m map because it is the SPIRE map with the highest resolution and the signal-to-noise ratio is the highest for most sources. However, this is not always the case for sources with the reddest continua, such as blazars. In fact, some of those with $S_{500\mu\rm m}> 35\,$mJy are much weaker and barely detected at $250\,\mu$m. We have therefore increased the search radius to check whether we may have missed some blazars with anomalously large errors on $250\,\mu$m coordinates. We found two sources, J083949.3$+$010437 and J001941.9$-$303116, whose nominal positions are separated by $11.09''$ and $10.64''$, respectively, from those of the BZCAT blazars PKS\,0837+012 and PKS\,0017$-$307. Their brightest pixels at $500\,\mu$m\footnote{See \url{https://www.h-atlas.org/img_cut_all/ .}} correspond to the blazar positions and their H-ATLAS flux densities are consistent with the blazar multi-frequency photometry. Furthermore, positions based on the $250\,\mu$m images may be offset by the contributions of nearby WISE (Wide-field Infrared Survey Explorer) sources. We have therefore identified the two sources with the blazars and included them in Table~\ref{tab:BZCAT}.

The continuum spectra from radio to ultraviolet of confirmed H-ATLAS blazars are shown in Figure~\ref{fig:blazarSEDs} (see Section~\ref{sect:radio_xmatch} for more details about points at radio and mm wavelengths).  The shapes of the observed SEDs in the selected range are remarkably similar (except for low-frequency excesses or self-absorption, to be discussed in Sect.~\ref{sect:blazarSEDs}) and overall smooth over about 7 orders of magnitude in frequency, despite variability (the data points are not simultaneous). The mean smoothed shape can be represented by a double power law:
\begin{equation}\label{eq:blazarSED}
S_\nu= A\left[(\frac{\nu}{\nu_0})^\alpha + (\frac{\nu}{\nu_0})^\beta\right]^{-1}
\end{equation}
with $\alpha=0.17$, $\beta=1.15$, $\nu_0=1000\,$GHz. The normalisation of the thick solid red line is $A=228\,$mJy. 
%The rest-frame SED still remain similar and smooth and can be described by a double power law with  $\alpha=0.10$, $\beta=1.25$, $\nu_0=2150\,$GHz, and $A=210\,$mJy. 

The shape of the average SED above several GHz is quite flat up to the knee at $\sim 1\,$THz (except for low-frequencies excesses or self-absorption, to be discussed in Sect.~\ref{sect:blazarSEDs}) and substantially steepens at higher frequencies. The differences of individual SEDs from the average one are limited, implying that flux density ratios, such as $S_{500\mu\rm m}/S_{1.4\rm GHz}$ and \textit{Herschel} colours, span a limited range of values in spite of the broadening of the ranges by variability in the case of non-simultaneous measurements.

%The uniformity of the average SED means that blazars span a limited range of flux density ratios over about 7 orders of magnitude in frequency, although variability may substantially broaden such range in the case of non-simultaneous measurements. We are interested, in particular, on the $500\,\mu$m to $1.4\,$GHz ratios. 
Also important, for our purposes, is that their sub-mm spectral slopes are different from those of dusty galaxies, in the sense of being much redder than those of the dustiest galaxies detected by \textit{Herschel} extragalactic surveys \citep[see Fig.~4 of][]{Negrello2017lensed}.

On account of the weakness of the blazar flux density at $250\,\mu$m, pointed out above, we extended the search radius for counterparts in $1.4\,$GHz catalogues to $15\,$arcsec, carefully checking the reliability of associations beyond $10\,$arcsec. For the equatorial and NGP fields we adopted the $1.4\,$GHz peak flux densities from the FIRST catalogue. For the SGP field, not covered by the FIRST survey, we used those from the NRAO VLA Sky Survey \citep[NVSS;][]{Condon1998}.

Given the low significance of the flux densities measured with the PACS instrument, we considered only the SPIRE data to define our selection criteria. We have computed the $S_{250\mu\rm m}/S_{350\mu\rm m}$, $S_{350\mu\rm m}/S_{500\mu\rm m}$ and $S_{500\mu\rm m}/S_{1.4\rm GHz}$ flux density ratios for all blazars in Table~\ref{tab:BZCAT}. The ratios span the ranges reported in Table~\ref{tab:ratios}. Preliminary ATCA follow-up observations (see Section~\ref{sect:ATCA_obs}) have uncovered 5 additional sources behaving as blazars (more details in section \ref{sect:ATCA_obs}). Their colours are within the ranges found for BZCAT blazars except for somewhat larger $S_{500\mu\rm m}/S_{1.4\rm GHz}$ and $S_{350\mu\rm m}/S_{500\mu\rm m}$  ratios ($\simeq 1.7$ and 1.55, respectively).

\begin{table}
    \centering

\caption{Flux density ratio ranges for known blazars (including those identified thanks to the first ATCA observations) in the different fields and final selection criteria for blazar candidates (last column).}
    \label{tab:ratios}
\begin{adjustbox}{width=\columnwidth}\begin{tabular}{cccc|c}
\hline    
Flux density ratios & equatorial & SGP & NGP &  final criteria\\
\hline
$ S_{250\mu\rm m}/S_{350\mu\rm m}$ & 0.6-1.15 & 0.73-1.19 & 0.58-1.0 & 0.5-1.2 \\
  $ S_{350\mu\rm m}/S_{500\mu\rm m}$ & 0.63-1.19 & 0.48-1.13 & 0.77-1.11 & 0.4-1.2\\
  $ S_{500\mu\rm m}/S_{1.4\rm GHz}$ & $<$0.75 & $<$1.61& $<$0.57 & $<$1.7\\
\hline    
    \end{tabular}
\end{adjustbox}

\end{table}

Based on these data, we have adopted the following selection criteria for blazar candidates, that avoid the risk of missing genuine blazars, at the cost of more work to clean the sample to remove non-blazars: $S_{500\mu\rm m}> 35\,$mJy; $0.5 < S_{250\mu\rm m}/S_{350\mu\rm m}< 1.2$, $0.4< S_{350\mu\rm m}/S_{500\mu\rm m}< 1.2$ and $S_{500\mu\rm m}/S_{1.4\rm GHz}< 1.7$.  

The selection criteria evolved through the years following the
updates of the H-ATLAS catalogue. As a consequence, different samples were selected for follow-up observations made in different epochs. Consistently with our  approach, we decided to include in Table\,\ref{tab:candidates} the 16 candidates with $S_{500\mu\rm m} \ge 35\,$mJy in the final H-ATLAS catalogues, selected throughout the
years, which, while not complying with the final criteria, are not far from being consistent with them. Such sources constitute a sort of ``control sample'' allowing us to check whether the final criteria were too restrictive in the sense of leaving out genuine blazars. We have flagged the 16 objects with an asterisk in all the tables. Among them, the most anomalous one is J002619.6$-$355344 whose final H-ATLAS flux density at $350\,\mu$m is only significant at $\sim 1.2\,\sigma$ and is incongruous with flux densities at the other two SPIRE wavelengths. It is in our control sample because data on the NASA/IPAC Extragalactic Database (NED) show that it has a flat spectrum between 1.4 and 4.8\,GHz and could therefore be a blazar missed by our selection. However, as discussed in the following, our radio measurements showed that this is not the case. Actually, none of the 16 objects turned out to be a blazar, implying that the new criteria increase the selection efficiency without losing genuine blazars.

In the following sections we will describe the follow-up observations and the cross-matches that allowed us to identify the genuine blazars.

\begin{table}
\caption{Setup of ATCA observations.}\label{tab:ATCAEpochs}
\hspace{-0.7cm}
\begin{adjustbox}{width=\columnwidth}
\begin{tabular}{ccccccc}
\hline
  Epoch & Project &
  Observing &
  Observed &
  Frequencies & ATCA & Resolution$^*$\\
  ID & ID &
   Dates&
  fields &
  [GHz] & Array& [arcsec$\times$arcsec]\\
\hline
A & C2673& 2012 Jul 14 & SGP  & 2.1 & H168 &64$\times$49\\
A & C2673& 2012 Jul 15 & SGP  & 5.5 - 9.0 & H168&46$\times$27\\
A & C2673& 2012 Jul 16 & SGP  & 33.0 - 39.0 & H168&27$\times$17\\
A & C2673& 2012 Jul 17 & Eq & 5.5 - 9.0,& H168&58$\times$26 - 36$\times$16\\
  &      &             &    &  33.0, 39.0 &                      &    8$\times$5 - 7$\times$4 \\
A & C2673& 2012 Jul 18 & Eq & 2.1 & H168 &101$\times$66\\
B & C2994& 2015 Jan 25 & SGP & 2.1 & 6A& no image\\
B & C2994& 2015 Jan 28 & SGP & 5.5 - 9.0 & 6A&no image\\
B & C2994& 2015 Jan 28 & Eq & 2.1, 5.5 - 9.0 & 6A&no image\\
C & C2673& 2020 Mar 28 & SGP & 2.1,& H168&$50\times35$,\\
  &      &             &    &  5.5 - 9.0 &                    &    $28\times19$ - $18\times12$ \\
  &      &             &    &  33.0 - 38.0 &                    &    $6\times4$ - $5\times3$ \\
\hline
\end{tabular}
\end{adjustbox}
*\,The reported values are the average of those obtained for the various images. They depend on the presence of antenna 6 and on the elevation of the sources at the time of observation.
\end{table}

\begin{table*}
%\begin{center}
\caption{ATCA flux densities, in mJy, for known and candidate blazars, listed in Tables~\ref{tab:BZCAT} and \ref{tab:candidates}, in the equatorial and SGP fields (upper and lower part of the table), observed in epoch A (2012). The coordinates, RA and Dec, are those of the radio peaks as detected in the highest resolution images. The listed errors represent the $1\,\sigma$ rms levels and do not include the calibration errors that can be estimated as $5\%$ of the flux densities and summed quadratically to the reported errors [see eq.~(\ref{eq:cal_err})]. Upper limits are estimated as $5\,\sigma$, $\sigma$ being the rms noise. Sources belonging to the control sample are flagged with an asterisk in the last column.
}\label{tab:ATCA2012}
\hspace{-0.8cm}
\setlength{\tabcolsep}{2pt}
\begin{tabular}{lccccccccccccl}
\hline
IAU ID & $S_{2.1}$ &$\sigma_{2.1}$ &$S_{5.5}$&$\sigma_{5.5}$ &$S_{9.0}$&$\sigma_{9.0}$ &$S_{33}$ &$\sigma_{33}$ &$S_{39}$&$\sigma_{39}$ & RA&Dec& class \\
\hline
J142631.1+014228  &	43.2&2.5&	13.6&0.2&	5.32&0.2&	0.57&0.11&	$<$0.8&	&	14:26:31.6&	+01:42:30.9&	cand*\\
J145146.1+010608  &	50.8&0.6&	106.4&0.4&	94.1&2.1&	105.5&2.1&	97.1&2.8&	14:51:45.9&	+01:06:19.4&	bl\\
\hline
J000348.6$-$344043  &	43.6&1.1&	23.9&1.2&	15.0&0.6&	4.8&0.1&	2.6&0.1&	00:03:48.5& 	$-$34:30:40.4&	cand*\\
J000438.2$-$321059  &	67.2&1.3&	46.0&0.3&	28.3&0.2&	5.5&0.1&	2.7&0.1&	00:04:38.3& 	$-$32:10:58.8&	cand*\\
J000935.6$-$321659  & 339.7&2.3&	293.9&1.0&	251.5&0.7&	194.4&0.6&	114.5&1.7&	00:09:35.5& 	$-$32:16:37.2&	bl\\
J001551.6$-$322927  &	19.4&0.3&	7.5&0.1&	3.5&0.5&	0.55&0.10&	$<$0.4&&	00:15:51.7&	$-$32:29:25.2&	cand\\
J003959.4$-$303333  &	26.8&1.3&	10.5&0.1&	5.1&0.1&	$<$0.5&	&	$<$0.4&&	00:40:00.0&	$-$30:33:36.4&	cand\\
J005802.3$-$323420  &	161.8&2.5&	220.8&5.5&	204.4&2.5&	155.1&1.0&	91.9&1.3&	00:58:02.2&	$-$32:34:20.8&	BZCAT\\
J013029.6$-$324847  &	21.7&0.8&	20.0&0.1&	17.1&0.5&	6.6&0.1&	3.5&0.1&	01:30:29.6&	$-$32:48:47.9&	bl\\
J014310.0$-$320056  &	120.1&2.4&	218.2&1.0&	252.6&1.6&	311.6&3.2&	215.0&2.2&	01:43:10.1&	$-$32:00:57.0&	BZCAT\\
J222321.6$-$313701  &	234.2&6.9&	383.3&1.2&	445.0&2.1&	516.0&4.3&	317.7&3.6&	22:23:21.6&	$-$31:37:02.5&	bl\\
J223454.5$-$325910  &	32.8&2.1&	21.4&0.2&	13.9&0.1&	2.9&0.1&	1.6&0.1&	22:34:54.7&	$-$32:59:07.1&	cand\\
J231448.5$-$313837  &	488.4&5.4&	615.2&1.6&	523.6&1.7&	334.7&1.7&	182.2&3.5&	23:14:48.5&	$-$31:38:39.8&	BZCAT\\
\hline
\end{tabular}     
%\end{center}
\end{table*}

\begin{table*}
%\begin{center}
\caption{ATCA flux densities, in mJy, for known and candidate blazars, listed in Tables~\ref{tab:BZCAT} and \ref{tab:candidates}, in the equatorial and SGP fields (upper and lower part of the table), observed in epoch B (2015). Flux densities in this epoch have been extracted from the visibilities exploiting the triple product technique. Listed errors represent the $1\,\sigma$ rms level and do not include the calibration errors that can be estimated as $5\%$ of the flux densities and summed quadratically to the reported errors  [see eq.~(\ref{eq:cal_err})]. Indications of extendedness are extracted from the visibilities as described in the text. The apex ''m'' on some flux density values indicates visibility patterns at that frequency that may be indicative of multiple sources in the field of view, while the ''e'' in the ''ext'' column flags sources that appear extended at any frequency. The pointing position of observations is reported: it can differ from the current H-ATLAS source position as they were defined on the basis of a previous version of the H-ATLAS catalogue. 
}\label{tab:ATCA2015}
\hspace{-0.8cm}
\setlength{\tabcolsep}{2pt}
\begin{tabular}{lccccccccccl}
\hline
IAU ID&$S_{2.1}$ &$\sigma_{2.1}$ &$S_{5.5}$&$\sigma_{5.5}$ &$S_{9.0}$&$\sigma_{9.0}$ &ext  & RA& Dec & class \\
\hline
J085419.0$-$003613  &	31.7&0.2&	10.2&0.3&	6.6&0.2 &e &  08:54:18.69 &  $-$00:36:19.14 &	cand*\\
J090910.1+012134  &	1627&0.3&	1925&0.3&	1853&0.2 & &09:09:10.10   &+01:21:35.96 &	BZCAT\\
J090940.2+020004  &	169.1&0.2&	149.1&0.4&	1282&0.3 & &09:09:39.85   &+02:00:05.33	&BZCAT\\
J113245.7+003427  &	389&0.3&	246.2&0.1&	200.5&0.1 && 11:32:45.63  & +00:34:27.71 &	BZCAT\\
J113302.8+001545  &	441&0.2&	350.6&0.1&	306.9&0.1 && 11:33:03.04  & +00:15:48.95& 	BZCAT\\
J113320.2+004053  &	534&0.2&	421&0.1&	379&0.1 && 	11:33:20.07   &+00:40:52.89 &BZCAT\\
J115043.6$-$002354  &	2440&0.3&	1510&0.2&	1190&0.1 &&  11:50:43.89  & $-$00:23:54.01 &	BZCAT\\
J120123.1+002830  &	413&0.2&	295&0.1&	261&0.1 &e & 12:01:23.20  & +00:28:29.01 &  	bl\\
J120323.5$-$015846  &	72&0.2&	22&0.1&	11.5&0.1 &e & 12:03:22.97  & $-$01:58:46.57 	&cand*\\
J120741.7$-$010636  &	124&0.3&	154&0.2&	212&0.1 && 12:07:41.65 &  $-$01:06:36.88	&BZCAT\\
J121008.3+011528  &	97&0.3&	40.6&0.1&	23.1&0.1 & &12:10:08.49   &+01:15:26.11	&cand\\
J121758.7$-$002946  &	247&0.2&	208&0.2&	241&0.1 &e & 12:17:58.68 &  $-$00:29:45.70 &	BZCAT\\
J121834.9$-$011954  &	236&0.3&	227&0.1&	220&0.1 && 12:18:34.95  & $-$01:19:54.76 	&BZCAT\\
J141015.9+005430  &	35.6&0.3&	15.9&0.2&	11.4&0.2 && 14:10:15.78  & +00:54:30.09 &	cand\\
J141004.7+020306  &	588&0.2&	485&0.1&	460&0.1 &e & 14:10:04.65  & +02:03:06.14 &	BZCAT\\
\hline
J001035.7$-$302745&	362&0.5&	333&0.1&	291&0.1 & & 00:10:35.61 &  $-$30:27:47.70  &	BZCAT\\
J010726.5$-$310822&	35.1$^m$&0.2&	5.0&0.1&	4.8&0.1 && 	01:07:26.96  & $-$31:08:29.00 &cand\\
J224041.0$-$335946  &	78$^m$&0.3&	32.2&0.3&	28.2&0.1 & &22:40:40.32  & $-$33:59:42.70&  	cand\\
J224500.0$-$343029  &	303&0.5&	121&0.2&	68.3&0.1 &&  22:45:00.28 &  $-$34:30:30.20 &	cand*\\
J224838.6$-$323551  &	646&0.1&	536&0.1&	455&0.1 &e &  22:48:38.54  &$-$32:35:51.40&	BZCAT\\
J231859.8$-$294738  &	19.1&0.2&	6.1&0.1&	7.0&0.1 & &  23:18:04.71   &$-$35:49:28.60 &	cand\\
J232125.5$-$315637  &	28.2&0.2&	4.9&0.1&	9.5&0.1 & &	 23:21:25.17  & $-$31:56:37.90&cand*\\
J232643.9$-$350420  &	27.1&0.2&	11.9&0.1&	7.9&0.1 & &	 23:26:44.22  & $-$35:04:18.00&cand\\
J234641.2$-$304858  &	31&0.2&	11.6&0.1&	8.5&0.1	 & &23:46:40.91 &  $-$30:49:07.20 &cand\\
J235347.4$-$303746  &	361&0.5&	415&0.1&	420&0.1 && 23:53:47.45  & $-$30:37:48.70   &	BZCAT\\
J235935.3$-$313343  &	710&0.1	&703&0.1&	606&0.1	 &&23:59:35.40  & $-$31:33:43.40& BZCAT\\

\hline
\end{tabular}     
%\end{center}
\end{table*}

\begin{table*}
%\begin{center}
\caption{ATCA flux densities, in mJy, for candidate blazars, listed in Table~\ref{tab:candidates}, in the equatorial and SGP fields (upper and lower part of the table), observed in epoch C (2020). The coordinates, RA and Dec, are those of the radio peaks as detected in the highest resolution images. The listed errors represent the detection $1\,\sigma$ rms level and do not include the calibration errors that can be estimated as $5\%$ of the flux densities and summed quadratically to the reported errors [see eq.~(\ref{eq:cal_err})]. Upper limits are estimated as 5 times the rms noise. }\label{tab:ATCA2020}
\hspace{-0.8cm}
\setlength{\tabcolsep}{2pt}
\begin{tabular}{lccccccccccccl}
\hline
IAU ID & $S_{2.1}$ &$\sigma_{2.1}$ &$S_{5.5}$&$\sigma_{5.5}$ &$S_{9.0}$&$\sigma_{9.0}$ &$S_{33}$ &$\sigma_{33}$ &$S_{39}$&$\sigma_{39}$ & RA&Dec& class \\
\hline
J000348.6$-$343043 &  43.90& 4.10 & 22.30 & 1.55&19.80 & 1.07 & 3.72 & 0.23 &  3.50 & 0.22 & 00:03:48.5& $-$34:30:40.8 &cand*\\
J000438.2$-$321059   & 71.00 & 3.78& 41.30 & 2.08 & 33.80 & 1.70 & 4.30 & 0.26 &  3.57 & 0.21 & 00:04:38.4& $-$32:10:58.9 &cand\\
J001551.6$-$322927   & 20.40 & 1.28& 7.53 & 0.39&  4.40 & 0.24 & $<0.6$  & ... &  $<0.5$  & ... & 00:15:51.6& $-$32:29:28.2 &cand*\\
J002213.1$-$302034    & 23.70 & 2.16& 13.30 & 0.69 & 9.85 & 0.51   &  1.80 & 0.14 &  1.80 & 0.69 & 00:22:12.8& $-$30:20:33.3 &cand\\
J002619.6$-$355344  & 90.40 & 5.63& 74.20 & 3.74 & 63.80 & 3.19 & 16.70 & 0.84 &  14.40 & 0.73 & 00:26:20.1& $-$35:53:35.8 &cand*\\
J003246.2$-$293109   & 343.00 & 17.66&115.00 & 5.89 & 64.60 & 3.26& 4.18 & 0.53 &  2.41 & 0.17 & 00:32:46.2& $-$29:31:10.1 &cand\\
J003959.4$-$303333   & 3.58 & 2.39& 10.40 & 0.61& 6.20 & 0.33 &  3.83 & 0.24 &$<0.8$ &  ...  & 00:40:00.4& $-$30:33:32.6 &cand*\\
J010726.5$-$310822   & 24.30 & 1.76 & 1.24 & 0.13& 8.94 & 0.45 & 1.13 & 0.06  &  0.68 & 0.10 & 01:07:27.2& $-$31:08:25.1 &cand\\
J011721.2$-$310845    & 81.50 & 5.97 & 99.40 & 5.64 & 119.00 & 6.53  & 146.00 & 7.34  &  146.00 & 7.37 & 01:17:21.4& $-$31:08:49.3 &cand\\
J013029.6$-$324847    & 27.20 & 2.35 & 12.40 & 0.64 & 22.20 & 1.11  & 5.39 & 0.77  &  5.30 & 0.35 & 01:30:29.7& $-$32:48:47.6 &cand\\
J222321.6$-$313701    &  295.00 & 14.76 &  357.00 & 17.87  &  540.00 & 27.13  &  569.00 & 29.29  &  572.00 & 28.81 & 22:23:21.6& $-$31:37:02.2 & bl\\
J223044.9$-$290038    & 44.10 & 2.28 & 19.20 & 0.96 & 14.30 & 0.73 & 2.04 & 0.14 &  1.97 & 0.20 & 22:30:45.6& $-$29:00:41.5 &cand\\
J223454.5$-$325910    & 43.40 & 2.39 & 22.20 & 1.12 & 19.10 & 4.67 & 4.89 & 0.28 &  4.59 & 0.27 & 22:34:54.6& $-$32:59:08.5 &cand\\
J224041.0$-$335946    & 49.50 & 2.74 & 33.80 & 1.72 & 34.50 & 1.75 & 11.40 & 0.58 &  9.72 & 0.51 & 22:40:39.7& $-$33:59:05.6 &cand\\
J224500.0$-$343029    & 298.00 & 14.95& 118.00 & 5.92&  81.80 & 4.13  &  11.00 & 0.57  &  9.32 & 0.52 & 22:45:00.3& $-$34:30:30.3 &cand*\\
J225036.6$-$323909    &  $<3.4$ & ... &  8.96 & 0.50  & 6.49 & 0.34  & 1.16 & 0.09  &  0.24 & 0.09 & 22:50:36.2& $-$32:39:02.7 &cand\\
J231859.8$-$294738    & 67.40 & 3.41 & 41.90 & 2.10 & 39.10 & 1.97& 12.20 & 0.64 &  11.00 & 0.58 & 23:18:59.9& $-$29:47:37.2  &cand\\
J232125.5$-$315637    & 19.10 & 1.19 & 5.56 & 0.32 & 3.96 & 0.21  & 0.40 & 0.08 &  $<0.6$ &...  & 23:21:24.9& $-$31:56:38.9 &cand*\\
J232643.9$-$350420    & 26.30 & 1.61 & 10.90 & 0.55 & 8.10 & 0.41 & 1.34 & 0.13 & $<0.7$& ... & 23:26:44.1& $-$35:04:21.5 &cand\\
J234641.2$-$304858    & 32.40 & 1.85 & 12.00 & 0.76 & 8.30 & 0.43 & 1.17 & 0.09&  0.99 & 0.11 & 23:46:41.0& $-$30:49:07.1 &cand\\
J234934.3$-$295254    & 56.70 & 3.29 & 19.20 & 1.05 & 13.60 & 0.75 & 1.43 & 0.20 &  1.86 & 0.25 & 23:49:33.9& $-$29:52:59.8 &cand\\
J235009.9$-$293948    & 44.60 & 2.72 & 25.10 & 1.33 & 17.90 & 0.90 & 2.86 & 0.26 &  2.60 & 0.13 & 23:59:09.1& $-$29:39:50.3 &cand*\\
\hline
\end{tabular}     
%\end{center}
\end{table*}

\begin{table*}
\begin{center}
\caption{VLA flux densities of 3 candidate blazars in the H-ATLAS equatorial fields. Peak ($S_p$) and total ($S_t$) flux densities are in mJy/beam and in mJy, respectively.}\label{TableVLA}
%\resizebox{\textwidth}{!}{
\setlength{\tabcolsep}{2pt}
%\begin{tabular}{|l|r|r|r|r|r|r|r|r|r|r|r|r|r|r|r|r|r|} % File blazar_cand_SGP_NVSS.tex
\begin{tabular}{lccccccccccccc} % File blazar_cand_SGP_NVSS.tex
\hline
IAU ID & $S_{p,6.2}$ & $E_{p,6.2}$ & $S_{t,6.2}$ &$E_{t,6.2}$ & $S_{p,22}$ & $E_{p,22}$ & $S_{t,22}$ &$E_{t,22}$ & $S_{p,33}$ &$E_{p,33}$ & $S_{t,33}$& $E_{t,33}$ & class \\
\hline
J085419.0$-$003613 &   0.57 &  0.01  &  20.22 &  0.40  &  0.43 &  0.01  &  3.70 &  0.07  &  0.33 &  0.01  &  0.33 &  0.01 & cand* \\
J120323.5$-$015846 &   17.07 &  0.34  &  35.76 &  0.72  &  1.57 &  0.03  &  5.47 &  0.11  &  0.75 &  0.02  &  2.49 &  0.05 & cand* \\
J121008.3+011528 &   35.14 &  0.70  &  35.14 &  0.70  &  5.93 &  0.12  &  5.93 &  0.12  &  3.26 &  0.07  &  3.26 & 0.07 & cand* \\
\hline
\end{tabular}          
\end{center}
\end{table*}

\section{Blazar identification in the radio-mm regime}\label{sect:radio_obs}

\subsection{ATCA observations}\label{sect:ATCA_obs}

We have refined our selection criteria by combining indications provided by existing blazar compilations and by three radio follow-up campaigns of selected candidates with the ATCA (Proposal IDs: C2673, C2994; P.I.: Massardi).

A summary of the details on the three observational epochs is presented in Table~\ref{tab:ATCAEpochs}, including observing dates, frequency ranges and resolutions obtained in the images. In this section we describe the observational setup and the data reduction for each of them. The extracted flux densities for the known and candidate blazars listed in Tables~\ref{tab:BZCAT} and \ref{tab:candidates} are reported in Tables~\ref{tab:ATCA2012}, \ref{tab:ATCA2015} and \ref{tab:ATCA2020}, respectively, for each of the 2012, 2015 and 2020 epochs.

The first epoch (A) in July 2012 targeted 39 sources in the equatorial and SGP fields selected using the criterion by \cite{LopezCaniego2013}. Only 9 of them remained after applying the final selection criteria and 4 more belong to the control sample. 

Different observing sessions were dedicated to different frequency setups and declination ranges of the sources, as listed in Table~\ref{tab:ATCAEpochs}. 
The correlator CFB 1M-0.5k continuum setup allowed for each receiver setting $2\times 2\,$GHz-wide bands that in our 3 settings were centered respectively at 2.1-2.1 GHz (i.e. both bands were centered at the same frequency but only one band was considered in the reduction), 5.5-9.0 and 33.0-39.0 GHz.

For all frequencies the primary calibrator was PKS1934-638, the bandpass calibrator was a bright quasar (either the same PKS1934-638, PKS0823-500 or PKS1253-055, depending on the frequency and  scheduled time slot for the observations). Calibration was performed with a standard procedure exploiting the Multichannel Image Reconstruction, Image Analysis and Display \citep[MIRIAD,][]{sault1995ASPC} software: flagging was performed at all frequencies in an automatic way with the task PGFLAG. If necessary, an additional manual flagging was performed after the calibration exploiting UVFLAG (this was particularly the case for $2.1\,$GHz, due to important radio frequency interference contamination, up to $50\%$ of the bandwidth).

The observations were carried out with a hybrid H168 array, performing several pointings on each target that were thus observed for at least 3 minutes at 2 different hour angles: this provided a nice coverage of the uv domain and hence the possibility of generating images at all frequencies. The sensitivity was enough to get a $\ge 5\,\sigma$ detection of each source at almost all frequencies, as listed in Table~\ref{tab:ATCA2012}. The reported errors are the $1\,\sigma_{\nu}$  rms of the images and do not include the calibration errors estimated to be $\sim 5\%$ of the flux densities. The global error should be estimated as:
\begin{equation}\label{eq:cal_err}
\hbox{E}_{\rm \nu}=\sqrt{\hbox{$\sigma$}_{\nu}^2+(0.05\times \hbox{S}_\nu)^2}.
\end{equation}
The table reports also the position of the peak in the radio images with the highest angular resolution where a source detection is achieved (typically 39 GHz): in all cases they are within $<10\,$arcsec of the H-ATLAS position.

The second follow-up campaign (January 2015, epoch B) was performed with an extended East-West 6A configuration on a selection (still performed with a preliminary criterion) of 46 H-ATLAS targets. The final selection recovered 21 of them and 5 more are in the control sample.
Observations in the CFB 1M-0.5k continuum configuration were run with the $2\times 2\,$GHz setup at frequencies $2.1-2.1\,$GHz and $5.5-9.0\,$GHz. Flagging and calibration were performed as for epoch A.
However, because of the observing conditions, for most of the targets only one 3-minutes cut was efficiently observed at any of the frequencies. This fact, together with the non-hybrid array, did not allow us to recover an image suitable to extract flux densities. Therefore, the flux densities were extracted from the triple product amplitude\footnote{The amplitude of triple product is the geometric average of the visibility amplitudes in a baseline closure triangle.}. The detection level is obtained from the variance of the visibilities divided by the square root of their number. This is less sensitive to phase errors for faint objects, even if the source is not at the phase center, as might be the case for some of our targets. 
%We investigated the efficiency of our approach by comparing triple products and image-based flux density determination by exploiting our own data collected in epoch A and C. A summary of the discussion is in the Appendix.

Due to the lack of images we also lack a straightforward determination of the source positions and of their structure. By analyzing the distribution of the visibility amplitudes as a function of the distance from the phase center (i.e. amplitude vs uv-distance plot) we defined if the source is point-like in the phase center (constant distribution) or may be extended, far from the phase center, or there are multiple sources in the field of view.  In the case of multiple sources the flux density definition from the visibilities is highly uncertain without proper modeling of the source distribution, which is not possible because of our lack of information on them. Fortunately, this seems to be the case only for 2 sources at $2.1\,$GHz. In the case of extended sources the ratio between the flux densities estimated on the longest visibilities and on the shorter ones is smaller than 1 \citep[indicatively $<0.85$, see][]{mm2008MNRAS384,chhetri2013MNRAS434} and in these cases the triple product may underestimate the real flux density. We have flagged such cases in Table~\ref{tab:ATCA2015}. If the source is point-like the triple product is a valid estimation of the flux density even far from the phase center.

The last epoch (C), in 2020, targeted 22 of the 24\footnote{At the time of the observations, J000935.6$-$321639 and J002615.8$-$351247 had already been identified as blazars in the literature and hence were not scheduled for follow-up.} candidates blazars in the SGP field (including 7 sources in the control sample) as listed in Table~\ref{tab:ATCA2020}.  Observational setup, reduction approach, and flux density determination procedures were the same as in epoch A (with the only exception that the highest--frequency band was centered at 38 GHz instead of 39 GHz to avoid radio frequency interferences).

\subsection{VLA observations}\label{sect:VLA_obs}

While ATCA observations covered the SGP sample, a VLA proposal to carry out high radio frequency observations of the candidate blazars in the NGP and equatorial fields got $\simeq 1\,$h at Priority B in the A-configuration (Proposal ID: VLA/20B-024, P.I.: Bonato). Three equatorial sources were observed in bands C (4--8 GHz), X (8--12 GHz), K (18--26.5 GHz) and Ka (26.5--40 GHz) on 21 Nov 2020. The fields of view (full width at half power of the primary beam) for the C, K and Ka bands were $\simeq 6.7$, 1.9 and 1.3 arcmin, respectively; the angular resolutions (full width at half-power of the synthesized beam) were 0.33, 0.089 and 0.059 arcsec, respectively. The calibrators used in the observations were 3C286 for flux scaling, and the blazars [HB89]\,0906+015 (J090910+012136) and [HB89]\,1148-001 (J115044-002354) for phase calibration. We used the calibrated Measurement Set (MS) provided by the National Radio Astronomy Observatory (NRAO; calibration done with Common Astronomy Software Applications, CASA, version 5.6.2-3). For each source the frequency channels of each band were combined (averaged) to obtain images centered at 6.2\,GHz (band C), 22\,GHz (band K) and 33\,GHz (band Ka). The imaging and the flux density extraction were done manually (using the version of the CASA software mentioned above).

The results are shown in Table~\ref{TableVLA}. The quoted errors are the 2\% calibration uncertainty\footnote{\url{science.nrao.edu/facilities/vla/docs/manuals/oss2017A/performance/fdscale}}: the noise contribution is negligible.
All the observed targets were included in the control sample after the definition of the final criteria, as described in Section~\ref{sect:sample_selection}.

\begin{figure}
\begin{center}
\includegraphics[width=0.5\textwidth]{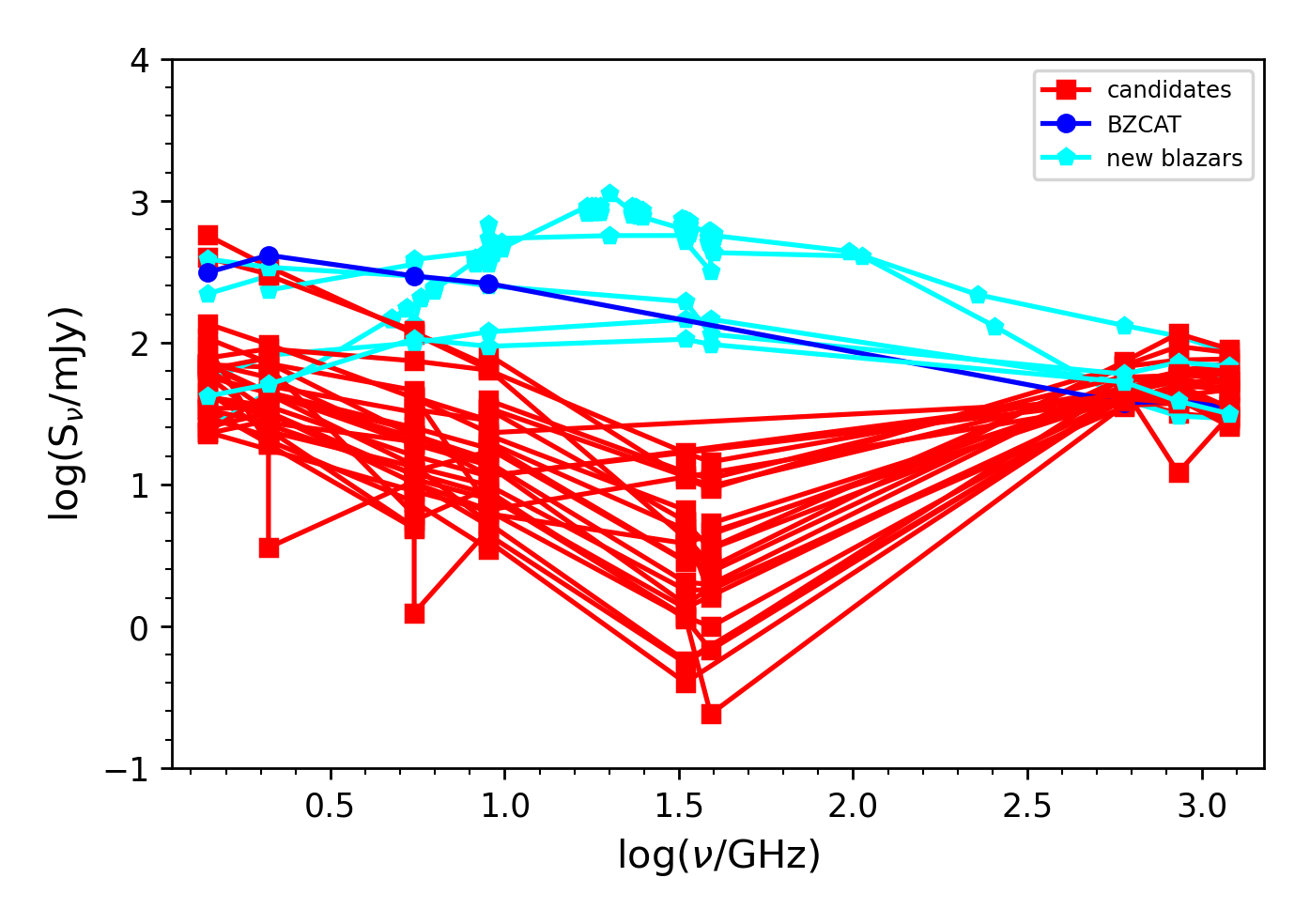}
\end{center}
\caption{Observed radio to sub-mm spectral energy distributions (SEDs) of candidate blazars in the H-ATLAS SGP field observed with the ATCA. We have highlighted in light blue the spectral behaviours similar to the mean SED of blazars already identified in the literature (dark blue line), and in red the behaviours ascribed to  steep-spectrum dusty galaxies with sub-mm emission dominated by thermal dust. Based on the spectral behaviour we identify with H-ATLAS sources 1 known blazar (J120123.1+002830) and 5 objects not previously recognized as genuine blazars. The data points include the NVSS flux densities at 1.4\,GHz, the ATCA measurements at 2.1, 5.5, 9.0, 33 and 38\,GHz, the H-ATLAS flux densities at 250, 350 and $500\,\mu$m, and  AT20G, PACO and ALMA flux densities when available (see the text for details).} 
\label{fig:SEDs}
\end{figure}

\subsection{Radio and mm cross-matches}\label{sect:radio_xmatch}
Our samples of known and candidate objects have been cross-matched with catalogues in the radio and mm bands to add more points to the SEDs in the region of interest in order to characterize the blazar spectral behaviour. Flux densities are included in Figs.\,\ref{fig:blazarSEDs} and \ref{fig:SEDs}, as well as in the SEDs of individual sources presented as Supplementary material (see Fig.~\ref{fig:example} for one example).

The Australia Telescope 20 GHz \citep[AT20G, ][]{mm2008MNRAS384, mm2011MNRAS412, murphy2010MNRAS402} survey covered the whole Southern sky at 20 GHz above a typical flux density limit of 40 mJy, with follow-up at 5.5 and 9.0 GHz. All the BZCAT sources with declination $<0^\circ$ in our selection have a counterpart in the AT20G catalogue. Only 4 sources among the candidates in the SGP (J000935.6$-$321639, J002615.8$-$351247, J11721.2$-$310845 and J222321.6$-$313701) have a counterpart in the AT20G. 

The \textit{Planck} ATCA Co-eval Observations \citep[PACO, ][]{mm2011MNRAS415, bonavera2011MNRAS416, bonaldi2013MNRAS428, galluzzi2017MNRAS465, Galluzzi2018} catalogue was built by following up with the ATCA a sample of AT20G sources brigther than 200 mJy in the right ascension interval 3--8 hr, simultaneously with the \textit{Planck} satellite, so that for each source at least 11 spectral points between 5.5 and 217 GHz were collected simultaneously and in several epochs.
Only 5 sources among the BZCAT blazars obeying our selection criteria and 2 sources among our candidates (J002615.8$-$351247, J222321.6$-$313701) have counterparts in at least one PACO observing epoch.

Finally, we cross-matched our selected sources with the ALMA calibrator catalogue\footnote{\url{https://almascience.eso.org/sc/}} to recover flux densities at frequencies above 100 GHz. The ALMA calibrator catalogue is a collection of ALMA observations of mm-band bright targets listed in various catalogues, including the BZCAT. Therefore almost all the equatorial and SGP sources included in Table~\ref{tab:BZCAT} have been observed at least once in one of the ALMA bands. The 2 candidates in Table~\ref{tab:candidates} that have a counterpart in PACO (J002615.8$-$351247 and J222321.6$-$313701) have a counterpart also in the ALMA calibrator catalogue.

%The flux densities of sources in the AT20G, PACO and ALMA calibrator catalogues are collected in tables in the Appendix.  
A search of the NED yielded the identification of two H-ATLAS sources brighter than 35\,mJy at $500\,\mu\rm m$ (J120123.1$+$002830 in the equatorial fields and J133108.4$+$303034 in the NGP field) with two known blazars not in BZCAT (the second blazar is the core of the bright radio source 3C\,286). 
The blazar FBQS\,J133037.6+250910 (the core of the bright radio source 3C\,287) is located on top of the brightest $350\,\mu$m and $500\,\mu$m pixels of H-ATLAS J133038.1+250901, while the $250\,\mu$m position is $11.16''$ away. We therefore concluded that also in this case the $500\,\mu$m flux density is dominated by the blazar.

\subsection{Radio to sub-mm SEDs: how to identify blazars}\label{sect:SED}
As mentioned in Sect.~\ref{sect:sample_selection}, Fig.~\ref{fig:blazarSEDs} collects the radio to UV SEDs of the BZCAT sources in our sample. Equation~(\ref{eq:blazarSED}), which describes the average SED of our sub-mm-selected blazars over 6 orders of magnitude in frequency, indicates that blazars SEDs in the range between 1 GHz to 1 THz are generally flat, although, below several GHz, some show either excess emission or evidence for self-absorption. We will return to that in Sect.\,\ref{sect:blazarSEDs}.

Figure~\ref{fig:SEDs} shows the radio to sub-mm SEDs of the sources selected in the SGP and equatorial fields, listed in Table~\ref{tab:candidates}. Two different behaviours can be identified in the frequency interval between $\sim 5$ and 600 GHz for this sample. On the one hand, a group of sources show a generally flat spectrum throughout the whole frequency range; the emission can be attributed to synchrotron all the way to sub-mm wavelengths, i.e. these sources can be classified as genuine blazars. 
On the other hand, the majority of sources show a spectrum with a minimum typically between 40 and 100 GHz. This behaviour can be described as radio emission fading with increasing frequency while the sub-mm emission is accounted for by thermal dust. This is the case for steep-spectrum radio sources hosted by dusty galaxies. 
In the figure, we distinguished with different colours the source identified as a blazar and the candidates that show the two behaviours.

\begin{figure}
%\begin{center}
\includegraphics[width=0.48\textwidth]{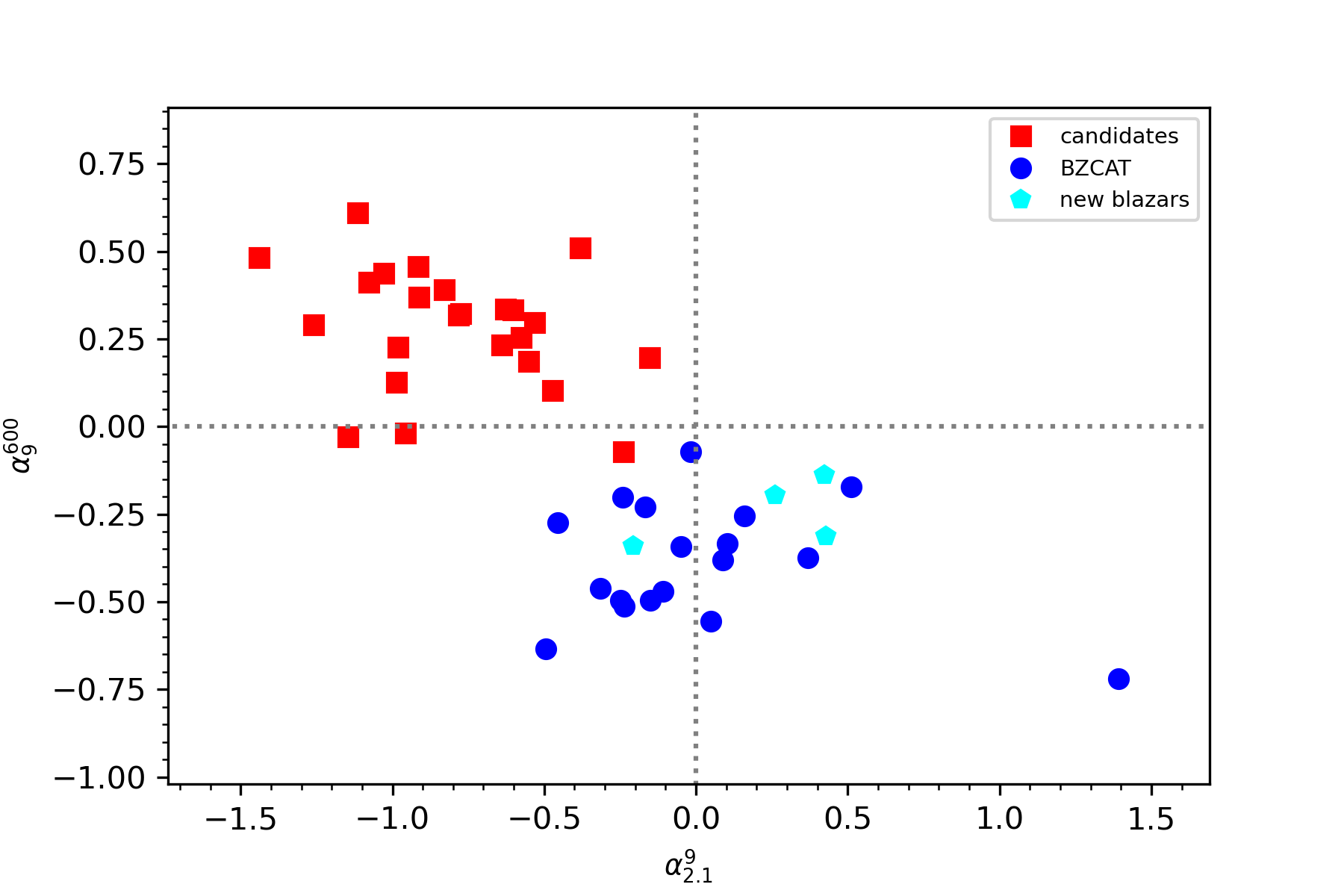} 
\includegraphics[width=0.48\textwidth]{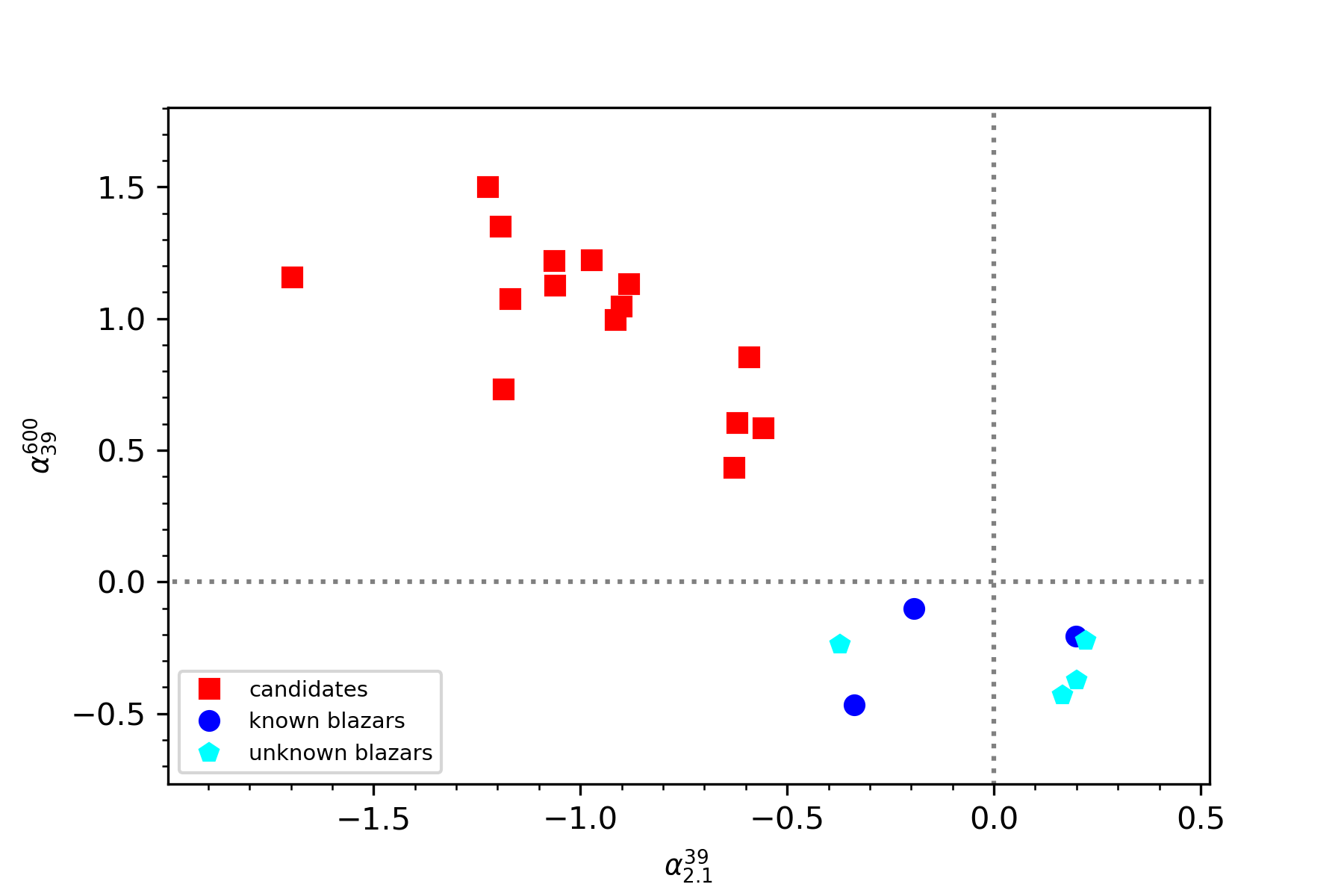}
\includegraphics[width=0.48\textwidth]{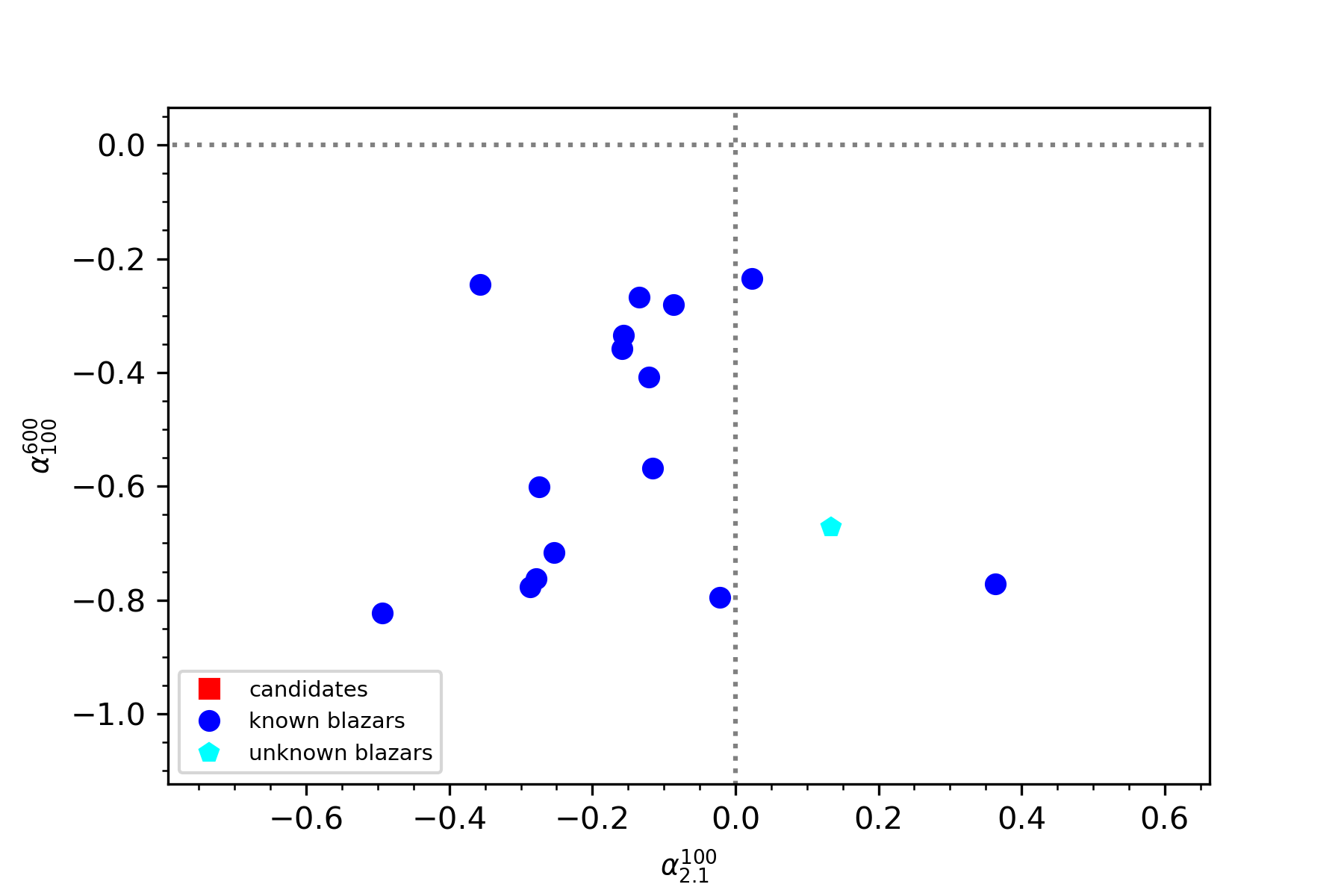}
%\end{center}
\caption{Radio/mm/sub-mm colour-colour plots for sources in our samples observed with the ATCA (upper and middle panels) and also included in the  ALMA calibrator catalogue (bottom panel). Error bars are in the worst case comparable to the symbol sizes.
%Spectral indices $\alpha$ between 2.1-x GHz and between x-600 GHz where x is 9, $\sim39$ and $\sim100$ respectively for panels a, b, and c for the sources for which we observed with ATCA and (for panel c) with a counterpart in the ALMA calibrator catalogue at $\sim 100$ GHz.
}
\label{fig:cc}
\end{figure}

We have further investigated the spectral differences among the two classes of sources by means of radio colour-colour plots. In Fig.\,\ref{fig:cc} we compared the spectral indices $\alpha_{\nu_1}^{\nu_2}$ (defined as $S_{\nu}\propto \nu^{\alpha}$ where $S_{\nu}$ is the flux density at frequency $\nu$) in different frequency ranges $\nu_1-\nu_2$ in GHz. 

The top panel shows $\alpha_{9}^{600}$ and $\alpha_{2.1}^{9}$ for sources we observed with the ATCA in epochs A and C. Seven SGP or equatorial candidates in Table~\ref{tab:candidates} have a flat spectrum in both frequency ranges. 

The middle panel shows, for the same sources, $\alpha_{39}^{600}$ versus $\alpha_{2.1}^{39}$ (note that in epoch C the reference frequency for the highest channel is 38 GHz instead of 39\,GHz). The plot clearly separates candidates in two groups with blazars located in the bottom right-hand corner, at $|\alpha_{2.1}^{39}|< 0.5$ and $-0.5<\alpha_{39}^{600}<0$.

The source J002619.6$-$355344 is classified as a flat-spectrum radio quasar on the basis of frequencies between 2.1 and 9 GHz. Our observations demonstrated that it steepens between 9 and 39 GHz, so that its spectrum is no longer flat all the way to sub-mm wavelengths. It is thus likely that its sub-mm emission is dust-dominated. Hence we did not include it in our sample of sub-mm blazars. This example clearly indicates that, while the selection must include the $\sim 2\,$GHz frequency point to be comprehensive, photometry at tens of GHz is crucial for the identification of genuine sub-mm blazars. 

In summary, only 6 targets (J120123.1+002830,  J145146.1+010608, J000935.6$-$321639, J011721.2$-$310845 J002615.8$-$351247, J002619.6$-$355344 and J222321.6$-$313701) among the candidates in Table~ \ref{tab:candidates} in the SGP and equatorial fields can be classified as genuine sub-mm blazars as their SEDs are consistent with being synchrotron-dominated from radio to sub-mm wavelengths. One of them (J120123.1+002830) was already identified in the literature as a blazar (see Section~\ref{sect:radio_xmatch}). J222321.6$-$313701 is listed in a catalogue of candidate blazars by Healey et al. (2008) and our spectral analysis confirms the classification. J145146.1+010608 was already classified as a candidate blazar by the previous analysis by our group \citep{LopezCaniego2013} and the blazar classification is confirmed by the more extensive radio follow-up presented here. The remaining 3 blazars (J000935.6$-$321639, J002615.8$-$351247 and J011721.2$-$310845) are new identifications.

In the bottom panel of Figure~\ref{fig:cc} we plotted the spectral indices $\alpha$ between 100 and 600\,GHz against those between 2.1 and 100 GHz for sources observed with the ATCA having a counterpart in the ALMA calibrator catalogue in Band 3 ($\sim 100\,$GHz).  %The ALMA calibrator catalogue comprises bright mm-wave targets selected from  other catalogues such as the AT20G and BZCAT. 
This plot confirms that blazar spectra are flat up to 100\,GHz and, on average, slightly steepen at higher frequencies. We can discern an indication of a slight hump between 40 and 100\,GHz.

% in the radio domain and flat or or slightly  downturning above 100GHz with no significant evidence of presence of dust. It indicates that the spectra appear flat at least up to 100 GHz, while there is a small trend towards steeper spectra above this frequency

\begin{figure}
\begin{center}
\includegraphics[width=0.48\textwidth]{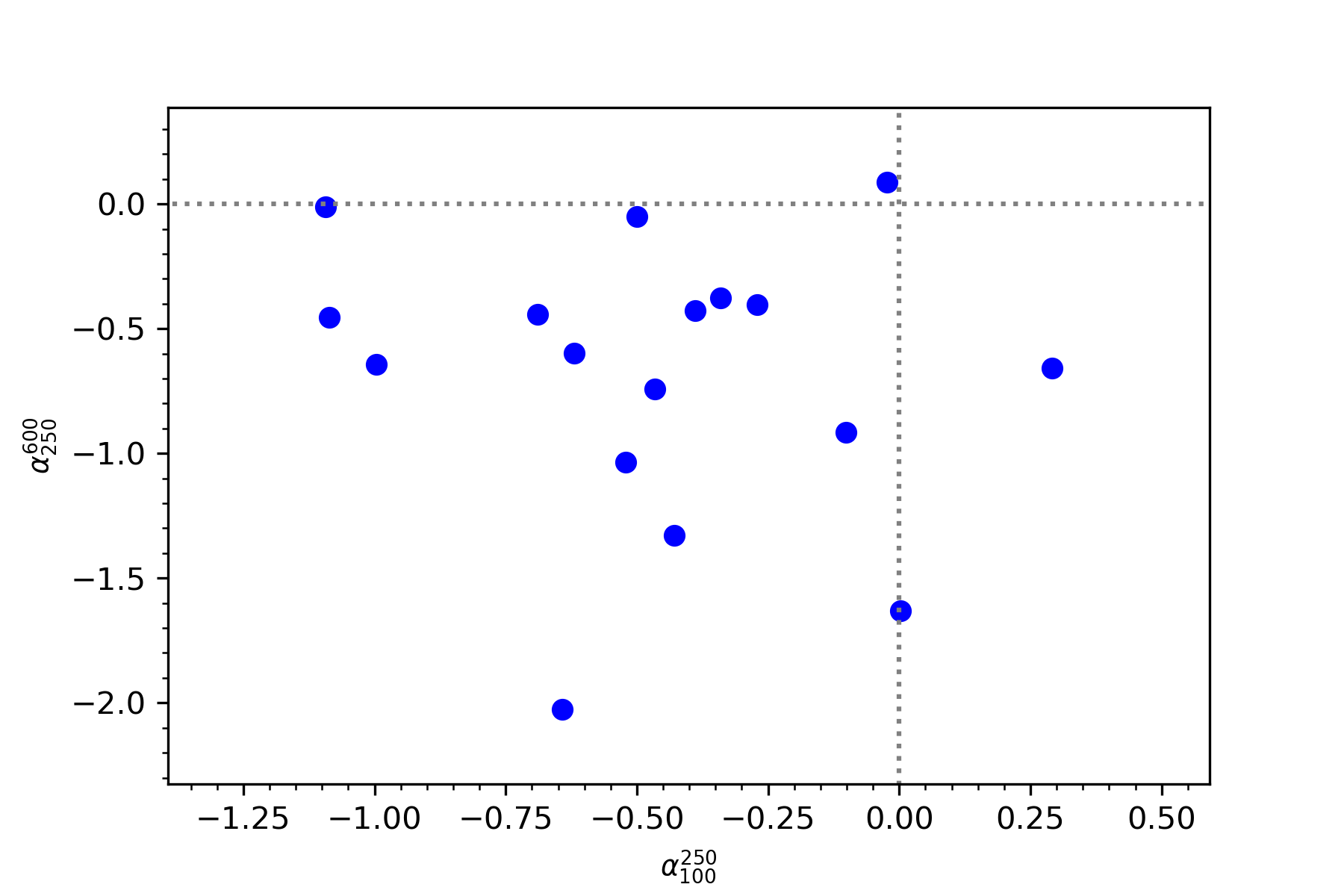}
\end{center}
\caption{Spectral indices $\alpha$ between 250 and 600\,GHz versus those between 100 and 250 GHz for BZCAT blazars in our sample with ALMA flux densities at $\sim 100$ and $\sim250$ GHz. Error bars are smaller than the symbol sizes.}
\label{fig:cc250}
\end{figure}

We further investigated the spectral behaviour in the mm/sub-mm domain in Figure~\ref{fig:cc250} by comparing the spectral indices in the ranges 100--250\,GHz and 250--600\,GHz for BZCAT blazars in our sample with ALMA flux densities at 100 and 250 GHz. The previously noted indication of a moderate steepening above 100\,GHz is confirmed and there is evidence of a further steepening of the average spectrum above 250\,GHz. 

The spectral data on the three sources we observed with the VLA indicate that they are steep--spectrum. The morphological information provided by our high-resolution data confirms this conclusion for two of them: J120323.5$-$015846 shows core plus jet and knot and J085419.0$-$003613 shows core plus lobes at 6.2\,GHz but only the core is visible at the higher frequencies. J121008.3+011528 is point-like. Peak flux densities refer to the core except for that of J085419.0$-$003613 at 6.2\,GHz, which refers to the east lobe. The core flux density at 6.2\,GHz of the latter source is 0.33\,mJy: the core spectrum is remarkably flat.

Since we did not get time to observe the NGP blazar candidates (Table~\ref{tab:candidates}) we exploited the uniformity of sub-mm blazar SEDs (Figure~\ref{fig:blazarSEDs}) to take the maximum advantage of all the photometry available in the literature in order to classify them. Such data suggest that they are all steep-spectrum in the radio and dominated by dust emission in the sub-mm, with the only exceptions of J133038.1+250901 and J133108.4$+$303034, as described in the previous section.

Summing up, our selection criteria, together with the information from the radio band, allowed us to identify a sample of 39 (31 already known and 8 newly recognized) confirmed sub-mm blazars and to investigate their spectral behaviour.

\section{Blazar SEDs}\label{sect:blazarSEDs}

We collected all the data available in the repository of the Space Science Data Center \citep[SSDC\footnote{\url{https://tools.ssdc.asi.it/SED/}};][]{2011arXiv1103.0749S} of the  Italian Space Agency and in the NASA/IPAC Extragalactic Database (NED)\footnote{\url{https://ned.ipac.caltech.edu/}} for the known and the newly identified blazars. We have complemented them with the H-ATLAS data and with our radio observations, and built the source SEDs.

Table\,\ref{tab:peaks} lists our estimates of the synchrotron peak frequencies of blazars with measured redshift. Such frequencies were derived by fitting with a parabola, as usual, the part of the rest-frame SED, in terms of $\nu L_{\nu}$, likely dominated by the Doppler boosted synchrotron emission. In practice we fitted data in the rest-frame range $5\le [\nu_{\rm rest}/\rm GHz] \le 10^5$.

Below $\sim 5\,$GHz the emission is frequently dominated by extended, steep-spectrum components, generally interpreted as leftovers of earlier activity episodes \citep[e.g.,][]{Ghisellini2010, Ghisellini2013}. As \citet[][p. 128, Sect. 8.9.1]{Ghisellini2013} puts it: ``At very low frequencies we expect that the emission is always dominated by the lobe, for all viewing angles''; Fig.~8.11 on p. 129 of the same reference shows that such steep-spectrum component can dominate up to several GHz. The low-frequency excess is particularly conspicuous in the cases of 3C286 (J133108.4+303034) and 3C287 (J133038.1+250901); its contribution at 1.4 GHz makes the spectral index between 1.4 and 5 GHz, conventionally used to discriminate between flat- and steep-spectrum sources, somewhat steeper than the canonical boundary of $-0.5$. However, the data show that the synchrotron emission of these sources extends up to IR/optical
wavelengths, as is the case for blazars, and straightforwardly account, in particular, for the sub-mm H-ATLAS flux densities. In contrast, the continuum spectra of steep-spectrum radio sources sink down at frequencies above some tens of GHz, due to electron ageing \citep{Kellermann1966}. Further evidence of the blazar nature of these sources is the strong variability especially in the optical/UV and X-ray bands. Moreover, 3C286 was also detected in gamma-rays with a gamma-ray/synchrotron energy ratio in the blazar range \citep{FermiLAT2019_4LAC}.

Figure~\ref{fig:blazarSEDs} also shows evidence of self-absorption below $\sim 5\,$GHz in the SEDs of some BZCAT blazars, again indicating that low-frequency data should not be used in the fit of the synchrotron peak. In the $10^5$--$10^6\,$GHz range blazar SEDs are frequently dominated by the thermal emission from the accretion disk \citep{Giommi2012, Castignani2015}. At still higher frequencies the inverse Compton peak pops up; its study is outside the scope of the present paper.

\begin{figure*}
\begin{center}
\includegraphics[width=0.48\textwidth]{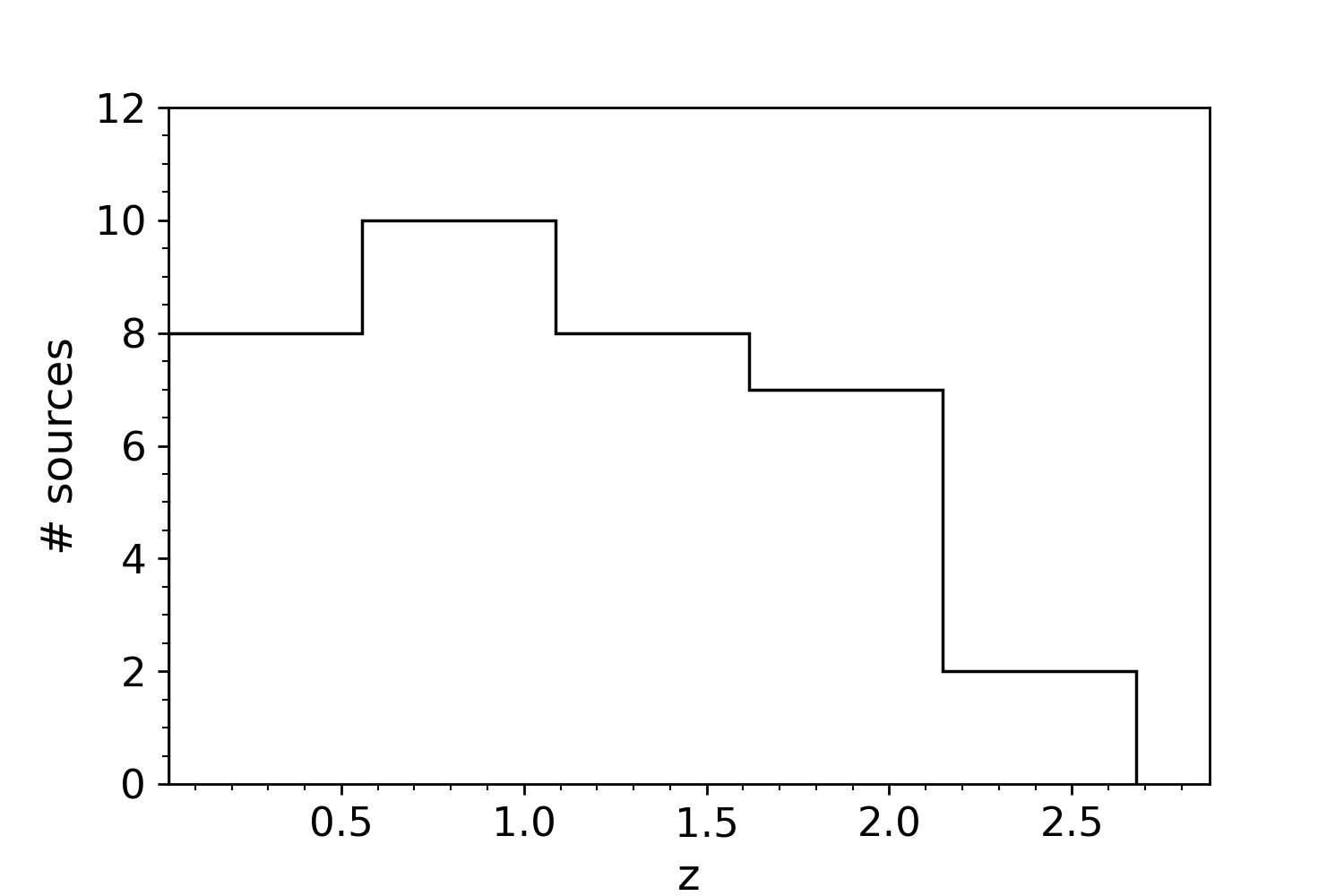}
\includegraphics[width=0.48\textwidth]{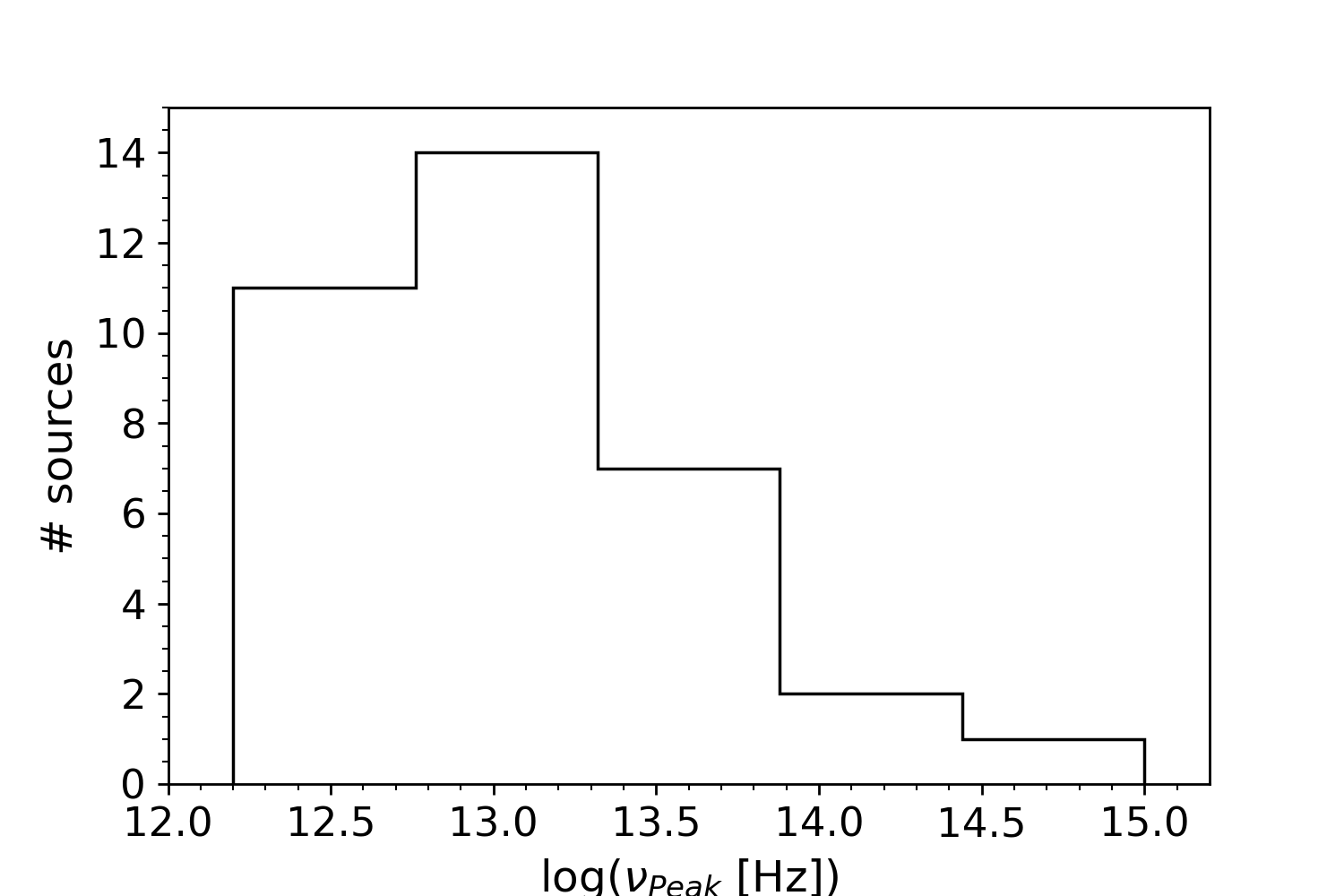}
\end{center}
\caption{Redshift (left) and rest-frame synchrotron peak frequency (right) distributions of the blazars in our sample. The peak frequencies are listed in Tab.\,\ref{tab:peaks}.}
\label{fig:peak}
\end{figure*}

The distribution of synchrotron peak frequencies is presented in the right-hand panel of  Fig.~\ref{fig:peak}; the median value is $\sim\mathbf{10^{13}}$\,Hz. The left-hand panel of the same figure shows the redshift distribution; redshifts range from 0.02 to 2.677, with a median value of 1.1.

\begin{table}
\caption{Synchrotron peak frequencies ($\nu_{\rm synch}$, THz), redshifts ($z$) and classifications of the blazars in our sample (U means ``uncertain type'' according to the BZCAT classification). The top part of the table contains the 31 BZCAT blazars (Table~\ref{tab:BZCAT}), the bottom part the 8 additional blazars in Table~\ref{tab:candidates}. The numbers in the first column are the order numbers in the respective tables.  }\label{tab:peaks}
\begin{tabular}{|l|l|r|r|r|}

\hline
   & HATLAS ID & z& $\nu_{\rm synch}$ & class \\
\hline

1  &J083949.3+010437    & 1.123   & 5.01 & FSRQ\\  
2  &J090910.1+012134    & 1.0245  & 5.01 & FSRQ\\   
3  &J090940.2+020004    & 1.7468  & 31.62 & BLLac\\   
4  &J113245.7+003427    & 1.2227  & 50.12 & BLLac\\     
5  &J113302.8+001545    & 1.1703  & 15.85 & FSRQ\\   
6  &J113320.2+004053    & 1.633   & 6.31 & FSRQ\\ 
7  &J115043.6$-$002354   & 1.9796  & 12.59 & FSRQ\\   
8  &J120741.7$-$010636   & 1.006   & 1.58 & FSRQ\\
9  &J121758.7$-$002946   & 0.4190  & 63.10 & BLLac\\  
10 & J121834.9$-$011954  & 0.1683  & 251.19 & BLLac\\   
11 & J141004.7+020306   & 0.1801  & 10.00 & BLLac\\     
12 & J001035.7$-$302745  & 1.190   & 19.95 & FSRQ\\  
13 & J001941.9$-$303116  & 2.677   & 5.01 & FSRQ\\   
14 & J003233.0$-$284921  & 0.324   & 63.10 & BLLac\\  
15 & J005802.3$-$323420  & $...$   & $...$ & BLLac\\ 
16 & J014310.0$-$320056  & 0.375   & 10.00 & U\\ 
17 & J014503.4$-$273333  &  1.148  & 15.85 & FSRQ\\   
18 & J224838.6$-$323551  &  2.268  & 5.01 & FSRQ\\
19 & J231448.5$-$313837  &  1.323  & 15.85 & FSRQ\\  
20 & J235347.4$-$303746  &  0.737  & 50.12 & BLLac\\  
21 & J235935.3$-$313343  &  0.990  & 5.01 & FSRQ\\   
22 & J125757.3+322930   &  0.8052 & 31.62 & FSRQ\\
23 & J130129.0+333703   &  1.0084 & 3.98 & FSRQ\\
24 & J131028.7+322044   &  0.9959 & 12.59 & U\\  
25 & J131059.2+323331   & 1.6391  & 12.59 & FSRQ\\   
26 & J131443.7+234828   & 0.2256  & 1000.00 & BLLac\\    
27 & J131736.4+342518   & 1.0542  & 10.00 & FSRQ\\   
28 & J132248.0+321607   & 1.3876  & 3.16 & BLLac\\   
29 & J132952.9+315410   & $...$   & $...$ & BLLac\\  
30 & J133307.4+272518   & 2.126   & 6.31 & FSRQ\\
31 & J134208.4+270933   & 1.1895  & 10.00 & FSRQ\\  
\hline                                
1  &  J120123.1+002830  & $...$   & $...$ &  ... \\
3  &  J145146.1+010608  & 2.115   & 6.31 & FSRQ \\ 
5 &  J000935.6$-$321639& 0.02564 & 199.53 & FSRQ \\ 
7 &  J002615.8$-$351247& 1.996   & 3.98 & FSRQ \\ 
10 &  J011721.2$-$310845& 0.106   & 39.81 &  ... \\
12 &  J222321.6$-$313701& $...$   & $...$ &  ... \\
25  & J133038.1+250901  & 1.055   & 3.16 & 3C287\\  
26  & J133108.4+303034  & 0.849886& 3.98 & 3C286\\  
\hline
\end{tabular} 
\end{table}

\begin{table}
\caption{H-ATLAS  equatorial, SGP and NGP (upper, central and lower part of the table, respectively) 
blazars with $\gamma$-ray associations within positional errors in the 4FGL catalogue \citep{FermiLAT2019_4LAC}.
SeMa (semi-major): long radius of error ellipse at 95\% confidence (deg.). %Flux and $\sigma_{\rm F}$: integral photon flux
%from 1 to 100\,GeV and its $1\,\sigma$ error (photons\,$\hbox{cm}^{-2}\,\hbox{s}^{-1}$).
Flux$_{\rm E}$ and $\sigma_{\rm E}$:
energy flux from 100 MeV to 100 GeV obtained by spectral fitting and its $1\,\sigma$ error (erg\,$\hbox{cm}^{-2}\,\hbox{s}^{-1}$).
Sep: angular separation (deg.) between the H-ATLAS and the 4FGL positions. }\label{tab:4FGL}
\setlength{\tabcolsep}{2pt}
\hspace{-0.6cm}
\resizebox{0.5\textwidth}{!}{
\begin{tabular}{|l|l|r|r|r|r|r|r|l|}
\hline
  \multicolumn{1}{|c|}{HATLAS ID} &
  \multicolumn{1}{c|}{4FGL ID} &
  \multicolumn{1}{c|}{SeMa} &
%  \multicolumn{1}{c|}{Flux} &
%  \multicolumn{1}{c|}{$\sigma_{\rm F}$} &
  \multicolumn{1}{c|}{Flux$_{\rm E}$} &
  \multicolumn{1}{c|}{$\sigma_{\rm E}$} &
  \multicolumn{1}{c|}{Sep} \\
\hline
J083949.3+010437 & J0839.8+0105 & 0.0638 &  6.25E-12 & 6.7E-13 & 0.0131 \\
J090910.1+012134 & J0909.1+0121 & 0.0224 &  3.24E-11 & 1.2E-12 & 0.0059 \\  % BZCAT
J090940.2+020004 & J0909.6+0159 & 0.0324 &  9.11E-12 & 8.5E-13 & 0.0104 \\  % BZCAT
J113245.7+003427 & J1132.7+0034 & 0.0186 &  2.53E-11 & 9.4E-13 & 0.0056 \\  % BZCAT
J120741.7$-$010636 & J1207.7$-$0106 & 0.0582 &  4.31E-12 & 5.2E-13 & 0.0054 \\  % BZCAT
J121758.7$-$002946 & J1218.0$-$0028 & 0.0441 &  6.08E-12 & 6.5E-13 & 0.0228 \\
J121834.9$-$011954 & J1218.5$-$0119 & 0.0259 &  1.20E-11 & 9.7E-13 & 0.0083 \\  % BZCAT
J141004.7+020306 & J1410.1+0202 & 0.0462 &  6.07E-12 & 5.5E-13 & 0.0188 \\
%
%J090910.1+012134 & J0909.1+0121 & 0.0224 & 2.47E-09 & 1.0E-10 & 3.24E-11 & 1.2E-12 & 0.0059 \\
%J090940.2+020004 & J0909.6+0159 & 0.0324 & 7.95E-10 & 7.0E-11 & 9.11E-12 & 8.5E-13 & 0.0104 \\
%J113245.7+003427 & J1132.7+0034 & 0.0186 & 2.25E-09 & 9.1E-11 & 2.53E-11 & 9.4E-13 & 0.0056 \\
%J120741.7$-$010636 & J1207.7$-$0106 & 0.0582 & 3.65E-10 & 4.8E-11 & 4.31E-12 & 5.2E-13 & 0.0054 \\
%J121834.9$-$011954 & J1218.5$-$0119 & 0.0259 & 1.48E-09 & 8.8E-11 & 1.20E-11 & 9.7E-13 & 0.0083 \\
%
\hline
J000935.6$-$321639 & J0009.7$-$3217 & 0.0567 &  2.10E-12 & 3.8E-13 & 0.0437 \\  % Candidate
J001035.7$-$302745 & J0010.6$-$3025 & 0.0552 &  4.52E-12 & 4.4E-13 & 0.0401 \\  % BZCAT
J003233.0$-$284921 & J0032.4$-$2849 & 0.1249 &  2.35E-12 & 3.7E-13 & 0.0262 \\  % BZCAT
J005802.3$-$323420 & J0058.0$-$3233 & 0.0241 &  2.12E-12 & 3.9E-13 & 0.0071 \\  % BZCAT
%J014310.0$-$320056 & J0143.5$-$3156 & 0.1476 &  1.45E-12 & 4.0E-13 & 0.1133 \\ % BZCAT source 63" away
J014503.4$-$273333 & J0145.0$-$2732 & 0.0446 &  1.21E-11 & 5.6E-13 & 0.0234 \\  % BZCAT
J224838.6$-$323551 & J2248.7$-$3235 & 0.0895 &  5.08E-12 & 5.2E-13 & 0.0274 \\  % BZCAT
J235347.4$-$303746 & J2353.7$-$3037 & 0.0578 &  4.66E-12 & 5.0E-13 & 0.0153 \\  % BZCAT
%
%J000935.6$-$321639 & J0009.7$-$3217 & 0.0567 & 1.83E-10 & 3.3E-11 & 2.10E-12 & 3.8E-13 & 0.0437 \\
%J001035.7$-$302745 & J0010.6$-$3025 & 0.0552 & 3.32E-10 & 4.0E-11 & 4.52E-12 & 4.4E-13 & 0.0401 \\
%J003233.0$-$284921 & J0032.4$-$2849 & 0.1249 & 1.94E-10 & 3.4E-11 & 2.35E-12 & 3.7E-13 & 0.0262 \\
%J005802.3$-$323420 & J0058.0$-$3233 & 0.0241 & 1.47E-09 & 7.7E-11 & 2.12E-12 & 3.9E-13 & 0.0071 \\
%J014310.0$-$320056 & J0143.5$-$3156 & 0.1476 & 1.00E-10 & 3.0E-11 & 1.45E-12 & 4.0E-13 & 0.1133 \\
%J014503.4$-$273333 & J0145.0$-$2732 & 0.0446 & 7.47E-10 & 5.5E-11 & 1.21E-11 & 5.6E-13 & 0.0234 \\
%J224838.6$-$323551 & J2248.7$-$3235 & 0.0895 & 2.10E-10 & 3.4E-11 & 5.08E-12 & 5.2E-13 & 0.0274 \\
%J235347.4$-$303746 & J2353.7$-$3037 & 0.0578 & 3.97E-10 & 4.5E-11 & 4.66E-12 & 5.0E-13 & 0.0153 \\
\hline
J125757.3+322930 & J1257.8+3228 & 0.0260 & 1.10E-11 & 7.6E-13 & 0.0239\\  % BZCAT
J130129.0+333703 & J1301.6+3336 & 0.0904 & 2.95E-12 & 5.0E-13 & 0.0396\\  % BZCAT
J131028.7+322044 & J1310.5+3221 & 0.0180 & 2.71E-11 & 2.0E-12 & 0.0141\\  % BZCAT
J131059.2+323331 & J1311.0+3233 & 0.0437 & 1.36E-11 & 1.7E-12 & 0.0112\\  % BZCAT
J131443.7+234828 & J1314.7+2348 & 0.0220 & 1.08E-11 & 8.4E-13 & 0.0070\\  % BZCAT
%J131615.9+301552 & J1316.5+3013 & 0.0736 & 1.08E-12 & 2.9E-13 & 0.0778\\  % MRK 785  z=0.04919
J131736.4+342518 & J1317.6+3428 & 0.0669 & 5.19E-12 & 4.9E-13 & 0.0459\\  % BZCAT
%J132301.1+294146 & J1323.0+2941 & 0.0190 & 1.51E-11 & 6.9E-13 & 0.0111\\  % Candidate
%J132545.5+350625 & J1326.0+3507 & 0.0895 & 1.09E-12 & 3.0E-13 & 0.0521\\  % IRAS F13234+3522 z=0.06401
J132952.9+315410 & J1330.4+3157 & 0.1580 & 1.76E-12 & 3.9E-13 & 0.1257\\  % BZCAT
J133108.4+303034 & J1331.0+3032 & 0.0750 & 2.23E-12 & 4.3E-13 & 0.0451\\  % BZCAT
J133307.4+272518 & J1333.2+2725 & 0.0466 & 6.00E-12 & 4.7E-13 & 0.0372\\  % BZCAT
\hline\end{tabular} }
\end{table}

\section{Comparison with other selections}\label{sect:comparisons}

\subsection{Near/mid-IR selection}

Also WISE near/mid-IR colours can be used to distinguish dusty galaxies from blazars \citep{MassaroF2011, DAbrusco2019}, as the decline of the dust emission is steeper (typically exponential) while that of the synchrotron blazar SED (see Fig. \ref{fig:blazarSEDs}) is close to a power-law.

To check the efficiency and the completeness of our selection, we have cross-matched H-ATLAS sources brighter than our flux-density threshold ($S_{500\mu{\rm m}}=35\,$mJy) with the two catalogues of radio-loud candidate blazars selected by \citet{DAbrusco2019} on the basis of WISE colours. The first catalogue (WIBRaLS2) includes radio-loud sources detected in all four WISE filters (effective wavelengths: 3.368, 4.618, 12.082 and $22.194\,\mu$m). The second catalogue (KDEBLLACS) contains radio-loud sources detected in the first three WISE passbands only. A search radius of $10''$ was used for our cross match.

We found 7 matches between the H-ATLAS equatorial sources and the WIBRaLS2 catalogue. The matches (J090910.1$+$012134, J113245.7$+$003427, J113302.8+001545, J115043.6$-$002354, J120123.1+002830, J121834.9$-$011954, J141004.7+020306) are all known blazars, confirming the validity of the selection. However, they are only $7/13=54\%$ of our blazar sample. The match with the KDEBLLACS catalogue yielded another
source (J114315.8-010733) which however has $S_{500\mu\rm m}/S_{1.4\rm GHz}=
4.56$, much larger than the maximum ratio of confirmed blazars. This object can
thus be classified as a dusty galaxy hosting a radio source, as confirmed by
its sub-mm colours.

As for SGP sources, we got 10 matches with WIBRaLS2: the 9 BZCAT blazars listed in Table~\ref{tab:BZCAT} plus J011721.2$-$310845, one of our candidates. Our ATCAobservations on this object allow us to classify it as a blazar. There are two matches with the KDEBLLACS catalogue (J014328.4$-$282148 and J224128.4$-$292820). Again, their ratios $S_{500\mu\rm m}/S_{1.4\rm GHz}$ (2.5 and 13.2, respectively) indicate that their sub-mm emission comes from thermal dust. %INTERPRETATION NOT
%SO CERTAIN FOR THE FIRST SOURCE (SPIRE COLOURS QUITE RED): ATCA OBSERVATIONS?

In the case of NGP sources we have 11 matches with WIBRaLS2. Nine are known
blazars (J125757.3+322930, J131028.7+322044, J131059.2+323331,
J131443.7+234828, J131736.4+342518, J132952.9+315410, J133108.4+303034,
J133307.4+272518, J134208.4+270933; only one known blazar is missed by the WIBRaLS2). One match (J125931.0+250449) is also selected as an NGP candidate
blazar by our criteria (Table \ref{tab:candidates}); its global SED indicates that its sub-mm emission is dust-dominated. The last match (J124831.9+271525) has $S_{250\mu\rm m}/S_{350\mu\rm m}=1.28$, exceeding the adopted upper limit ($S_{250\mu\rm m}/S_{350\mu\rm m}< 1.15$), again indicating a thermal dust dominated sub-mm emission. The only match with the KDEBLLACS catalogue (J133941.1+285544) has
$S_{500\mu\rm m}/S_{1.4\rm GHz}\simeq 3.5$, indicative of a radio source hosted by a dusty galaxy.

\subsection{Radio/optical selection}

\citet{Itoh2020} built a new catalogue of blazar candidates, named Blazar Radio and Optical Survey (BROS). The BROS includes 88,211 radio sources at $\hbox{Dec}> 40^\circ$, outside the Galactic plane region ($|b|>10^\circ$), with flat spectra ($\alpha > -0.6$) between 0.15 and 1.4\,GHz. Sources were selected cross-matching the Tata Institute of Fundamental Research GMRT (Giant Metrewave Radio Telescope) Sky Survey \citep[TGSS;][]{Intema2017} at 0.15 GHz with the NVSS catalogue. Optical identifications of radio sources were obtained using the Panoramic Survey Telescope and Rapid Response System (Pan-STARRS) photometric catalogue \citep{Chambers2016}.

A cross-match of the BROS with the H-ATLAS catalogues with a search radius of 10 arcsec yielded 21, 31 and 44 matches with $S_{500\mu\rm m}\ge 35\,$mJy in the equatorial, NGP and SGP fields, respectively.

Nine out of the 21 equatorial matches are blazars listed in Table~\ref{tab:BZCAT} (the missed known blazars are J083949.3+010437, J120123.1+002830 and J121758.7$-$002946); one (J121008.3+011528) is an equatorial source selected by our criteria (Table~\ref{tab:candidates}), shown by our VLA observations to be steep-spectrum (Table~\ref{TableVLA}). All the other 11 matches have sub-mm colours indicative of thermal dust emission.

The 31 NGP matches include 9 blazars listed in Table~\ref{tab:BZCAT} (missed known blazars: J131028.7+322044 and J132248.0+321607) and three of our NGP blazar candidates in Table~\ref{tab:candidates} (J125931.0+250449, J133038.1+250901 and J133940.4+303312). As mentioned above, J133038.1+250901 (the core of the bright radio source 3C\,287) is a blazar while neither of the other two sources can be classified as a sub-mm blazar. The sub-mm SEDs of 15 out of the other 20 matches have a thermal dust shape. Two of the remaining 5 sources (J125326.3+303641 and J134232.1+350713) have sub-mm SEDs consistent with those of the bright steep-spectrum radio sources B2\,1251+30 and 4C+35.30 located at 9 and 8 arcsec from the nominal H-ATLAS positions, respectively. The last two sources have insufficient data for a clear classification; however, their relatively large $S_{500\,\mu\rm m}/S_{1.4\rm GHz}$ ratios (2.8 for J133435.6+294958 and 3.2 for J134139.7+322834) suggest that the sub-mm SEDs are dust-dominated.

As for the SGP, we have 8 matches with blazars listed in Table~\ref{tab:BZCAT} (missed: J014310.0$-$320056 and J224838.6$-$323551) and 11 matches with our candidates in Table~\ref{tab:candidates}. Only one of them (J222321.6$-$313701) turned out to be a blazar, based on our ATCA observations. The literature data on the remaining 25 matches found in the NED are quite limited but we didn't find evidence of blazar SEDs.

\begin{figure*}
\begin{center}
\includegraphics[width=0.48\textwidth]{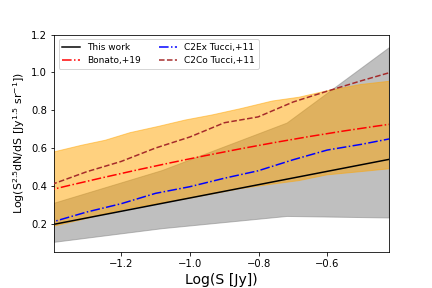}
\includegraphics[width=0.48\textwidth]{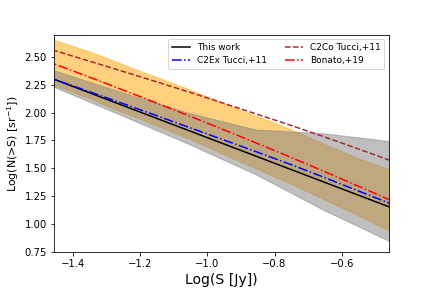}
\end{center}
\caption{Euclidean normalized (left) and integral (right) counts of blazars at $500\,\mu$m. The solid black lines show  our best fit, while the grey bands represent the Poisson 68\% confidence errors. Also plotted, for comparison, are the best-fit counts (dot-dashed red line) with their Poisson errors (orange band) obtained extrapolating the estimate obtained by \citet{Bonato2019} at 467\,GHz (ALMA band 8).  Poisson errors were computed following \citet{Gehrels1986}. The dot-dashed blue and the dashed brown lines show the predictions of models C2Ex and C2Co by \citet{Tucci2011} at 600 GHz ($500\,\mu$m), kindly provided by Marco Tucci. }\label{fig:counts}
\end{figure*}

\subsection{Gamma-ray selection}

As already mentioned, blazars dominate the extragalactic $\gamma$-ray sky. They constitute 98\% of  AGNs detected by the \textit{Fermi} Large Area Telescope (LAT). In turn, AGNs represent at least 79\% of the high Galactic latitude sources in the fourth \textit{Fermi}-LAT catalogue \citep[4FGL; cf.][]{FermiLAT2019_4LAC}. Table~\ref{tab:4FGL} shows the results of a cross-match of H-ATLAS sources with the 4FGL catalogue using as search radius the 95\% confidence semi-major axis of the 4FGL error ellipse.

The H-ATLAS sources selected for the cross-match were required to be brighter than $S_{500\mu\rm m}= 35\,$mJy and to have, within $15''$ a FIRST or NVSS counterpart with $S_{1.4\,\rm GHz}> 5\,$mJy. The latter flux density limit was imposed to get rid of galaxies whose radio emission could be powered by star formation. We purposely adopted looser criteria, compared to those adopted before, in order to check whether we had missed some true blazars. 

In the equatorial fields, we found 8 matches, listed in Table~\ref{tab:4FGL}. They are all known blazars, included in Table~\ref{tab:BZCAT}. One of them is the $\gamma$-ray source 4FGL\,J0839.8+0105, identified with the blazar PKS\,0837+012 which, in turn, we identify with the H-ATLAS source J083949.3+010437.

%The other 2 matches, J084237.3+025423 and J085941.4+004911, don't have any
%FIRST or NVSS counterpart within the H-ATLAS positional error box. Hence they
%are not plausible blazar candidates. Both matches are most likely random
%occurrences. 4FGL\,J0842.5+0251 is identified with the blazar
%NVSS\,J084225+025251 which  is $\simeq 3.3'$ apart from J084237.3+025423. 4FGL
%J0859.8+0053 has no identification in the 4FGL catalogue.  It lies at the
%border of the search radius.

Eight matches were found also in the SGP. We have discarded the match of H-ATLAS J234158.9$-$292116 ($S_{1.4\,\rm GHz}=19.3\pm 0.7\,$mJy) with 4FGL J2341.8-2917 because, in spite of the small separation ($3.9''$) it is probably a chance association. The most likely identification of the $\gamma$-ray source is the blazar PKS\,2338-295 which is located $6.6'$ away from J234158.9-292116 but well within the 4FGL error box. This conclusion is supported by the relatively blue colour of the H-ATLAS source, $S_{250\mu\rm m}/S_{350\mu\rm m}= 1.29$, above the limit $S_{250\mu\rm m}/S_{350\mu\rm m}< 1.15$. The first of the other 7 matches (Table~\ref{tab:4FGL}) is the newly recognized sub-mm blazar J000935.6$-$321639 (see Table~\ref{tab:candidates}). The remaining 6 matches are blazars listed in Table~\ref{tab:BZCAT}.

In the NGP we got 10 matches (Table~\ref{tab:4FGL}), 9 of which are listed in Table~\ref{tab:BZCAT}. The associations of the tenth source, J132301.1+294146, with the $\gamma$-ray source 4FGL J1323.0+2941 is doubtful. In \citet{FermiLAT2019_4LAC} the source is associated to the bright steep-spectrum radio source 4C\,+29.48 which is also well within the $\gamma$-ray error box. The spectral index between 1.4 and 4.8\,GHz of the radio source associated to J132301.1+294146 is $-0.66$, suggesting a steep-spectrum classification and its global SED is indicative of a dust-dominated sub-mm emission.

In conclusion, a large fraction of our blazars were recovered by the multiwavelength cross-matches we performed and no additional sub-mm blazars were found, indicating that our sample is 100\% complete.

\section{Number counts of sub-mm selected Blazars}\label{sec:counts}

The differential counts per dex estimated from our sample are listed in Table~\ref{table:counts} together with their Poisson errors, computed using the tables by \citet{Gehrels1986}. In the flux density range covered by our data the counts can be described by ${\rm d}N/{\rm d}S\simeq 5.6\,S^{-2.1}\,\hbox{Jy}^{-1}\,\hbox{sr}^{-1}$, with $S$ in Jy. We have defined the effective flux density, $S_{\rm eff}$, for each bin as the mean flux density weighted by the counts:
\[
S_{\rm eff} = \frac{\int_{S_1}^{S_2}\, ({\rm d}N/{\rm d}S)\,S\,{\rm d}S }{\int_{S_1}^{S_2}\, ({\rm d}N/{\rm d}S)\,{\rm d}S}.
\]
Figure~\ref{fig:counts} shows the best-fit blazar integral  and Euclidean normalized differential counts at $500\,\mu$m (right- and left-hand panel, respectively; solid black lines), with their Poisson uncertainties (grey bands). 

Our counts are in reasonably good agreement with, although somewhat lower than the estimate obtained by \citet{Bonato2019} at 467\,GHz; such estimate, extrapolated to 600\,GHz using the mean blazar SED of eq.~(\ref{eq:blazarSED}), is shown by the dot-dashed red lines with the orange error bands. Extrapolations of the estimate by \citet{Bonato2019} at 673\,GHz yielded essentially the same results. These estimates were derived using flux density measurements of ALMA calibrators. Counts were obtained convolving the South Pole Telescope (SPT) $95$\,GHz counts of radio sources  by \citet{Mocanu2013} with the distributions of flux density ratios between ALMA band 8 (385--500 GHz) or band 9 (602--720 GHz) and band 3 (84–116 GHz). 

The C2Ex model by \citet[][dot-dashed blue lines]{Tucci2011} yields counts at 600 GHz remarkably close to our best fit while their model C2Co (dashed brown lines) overpredicts the counts. The two models differ from each other in the size, $r_M$, of the synchrotron emitting region of FSRQs. This region is more extended, with $r_M$ in the range 0.3--10\,pc, in the C2Ex case and a factor of 10 more compact, with $r_M$ in the range 0.03--1\,pc, in the C2Co case. As a consequence, the synchrotron peak frequencies are typically lower by the same factor for the C2Ex model. The distribution of peak frequencies of BL Lacs is the same for both models and extends to higher frequencies than that of FSRQs.

\begin{table}
\begin{center}
\caption{Differential blazar counts, $\rm dn/d\log S$ ($\hbox{sr}^{-1}\,\hbox{dex}^{-1}$), at $500\,\mu$m. Flux densities are in Jy. $N$ (last column) is the number of blazars in the bin. Poisson 68\% confidence errors were computed using the tables of \citet{Gehrels1986}.}\label{table:counts}
\begin{tabular}{rrcr}
\hline
\multicolumn{1}{c}{Flux range}&\multicolumn{1}{c}{$S_{\rm eff}$} & $\rm dn/d\log S$ & $N$\\
\hline
$0.035-0.070$ &  $0.040$ &  $449 (+118,-96)$ &  22 \\
$0.070-0.139$ &  $0.083$ &  $196 (+80,-59)$ &  11 \\
$0.139-0.278$ &  $0.191$ &  $75 (+52,-34)$ &  5 \\
$0.278-0.555$ &  $0.379$ &  $34 (+47,-24)$ &  2 \\
\hline
\end{tabular}
\end{center}
\end{table}

\section{Conclusions}\label{sec:conclusions}

We have built a sample of 39 blazars with $S_{500\mu\rm m}>35\,$mJy extracted from the H-ATLAS survey catalogues, out of a total of
%9671 equatorial 9719 NGP, 16504 SGP
35,864 sources brighter than that flux density limit. To find out such a tiny minority ($\simeq 0.1\%$) of sources we have defined criteria reflecting the properties of known blazars detected by the H-ATLAS survey. First, they must have counterparts in large area radio surveys at 1.4\,GHz (FIRST for sources in H-ATLAS equatorial and NGP fields, NVSS for sources in the SGP field) with $S_{500\mu\rm m}/S_{1.4\rm GHz}<1.7$. Second, their sub-mm spectrum must be flat ($0.5 < S_{250\mu\rm m}/S_{350\mu\rm m}< 1.2$ and $0.4< S_{350\mu\rm m}/S_{500\mu\rm m}< 1.2$) with colours redder than those of the overwhelming majority of dusty galaxies.

The fact that the blazar sub-mm spectra are generally declining with increasing frequency implies that some of them are weak or only barely detected at $250\,\mu$m, the reference wavelength for the H-ATLAS astrometry, due to its better angular resolution (only local galaxies are detected by the H-ATLAS survey at the still shorter PACS wavelengths). As a result, 4 blazars were found to have nominal ($250\,\mu$m) positions offset from the radio position by slightly more than $10''$ but with $500\,\mu$m images consistent with the radio position and H-ATLAS photometry fully compatible with literature measurements of the blazar SEDs: they were accepted as valid identifications of the $500\,\mu$m detections.

Known sub-mm detected blazars, mostly from the BZCAT, were found to have remarkably similar spectral shapes from GHz to UV frequencies. Their typical spectral index is quite flat ($-0.17$) at low frequencies and gradually steepens up to $\simeq -1.15$ at high frequencies, with a knee at $\simeq 1\,$THz. An analytic representation of the typical spectrum is given by eq.~(\ref{eq:blazarSED}).

ATCA multi-frequency radio observations of candidate blazars in the equatorial and SGP fields were carried out, resulting in uncovering 6 blazars brighter than $S_{500\mu\rm m}=35\,$mJy. Of them, one was already classified as a genuine blazar in the literature, while the remaining 5 are confirmations of past indications or new classifications as blazars. The ATCA observations were complemented with VLA observations of three equatorial candidates, none of which turned out to be a blazar.

ATCA observations and counterparts in the ALMA calibrator catalogue allowed us to confirm that the blazar spectra are relatively flat from 1 to 600 GHz, with an indication of a slight hump between 40 and 100 GHz and a progressive steepening at higher frequencies. 

Our VLA proposal did not get time to observe candidates in the NGP H-ATLAS field. Our analysis of literature photometric data found only 2 of them with a SED clearly consistent with the typical blazar SED; both were previously classified as blazars.

To check whether our approach missed some blazars, we have also included in our analysis a control sample of 16 sources with $S_{500\mu\rm m}=35\,$mJy in the final HATLAS catalogue but with flux density ratios somewhat outside the ranges adopted for the final selection. None of them turned out to be a genuine blazar.

Furthermore, we have cross-matched H-ATLAS sources brighter than $S_{500\mu\rm m}=35\,$mJy with large catalogues of blazar candidates, namely the WIBRaLS2 and the KDEBLLACS catalogues, based on WISE colours \citep{DAbrusco2019}, and the BROS catalogue including radio sources with flat spectra between 0.15 and 1.4\,GHz \citep{Itoh2020}. %Search radii of up to 15 arcsec were used to avoid the risk of missing blazars with poor $250\,\mu$m astrometry. 
The cross matches recovered a large fraction (but not all) of our blazars. None of the other matches had continuum spectra indicative of being blazars or, at least, of having a synchrotron-dominated sub-mm emission. 

Also no additional sub-mm blazars were found in the cross-match with the fourth \textit{Fermi}-LAT catalogue \citep{FermiLAT2019_4LAC}, but 24 of blazars in our sample have a reliable $\gamma$-ray counterpart. We conclude that we did not find any indication of incompleteness of our sample.

The derived counts are compared with model predictions finding good consistency with the C2Ex model by \citet{Tucci2011} and with estimates by \citet{Bonato2019} based on ALMA data.

\section*{Acknowledgments}
We are grateful to Marco Tucci for having provided his model counts at 600\,GHz. We thank the anonymous referee for a very careful reading of the manuscript and stimulating comments. This research has made use of \textit{Herschel}-ATLAS data. The \textit{Herschel}-ATLAS is a project with \textit{Herschel}, which is an ESA space observatory with science instruments provided by European-led Principal Investigator consortia and with important participation from NASA.  We also made use of the TOPCAT software \citep{TOPCAT2005}, of the ASI SSDC repository and of the NED. The NED is operated by the Jet Propulsion Laboratory, California Institute of Technology, under contract with the National Aeronautics and Space Administration. 
This paper made use of the following archive ALMA data: ADS/JAO.ALMA\#2011.0.00001.CAL. ALMA is a partnership of ESO (representing its member states), NSF (USA) and NINS (Japan), together with NRC (Canada), MOST and ASIAA (Taiwan), and KASI (Republic of Korea), in cooperation with the Republic of Chile. The Joint ALMA Observatory is operated by ESO, AUI/NRAO and NAOJ.

MM and AL acknowledge the support from grant PRIN MIUR 2017 - 20173ML3WW\_001. MB acknowledges support from the Ministero degli Affari Esteri della Cooperazione Internazionale - Direzione Generale per la Promozione del Sistema Paese Progetto di Grande Rilevanza ZA18GR02. 
LB and JGN acknowledge the PGC 2018 project PGC2018-101948-B-I00 (MICINN/FEDER).
AL is supported by the EU H2020-MSCAITN-2019 Project 860744 ‘BiD4BEST: Big Data applications for Black hole Evolution STudies’.

\section*{Data availability}
All the catalogued data used for this paper are publicly available in the catalogues cited in the text.

The SEDs of individual sources used for our analysis are made available as online supplementary material.

ATCA and VLA data are publicly available through the ATCA and VLA archives.

\bibliographystyle{mnras}
\bibliography{HATLASblazar} % if your bibtex file is called example.bib

%\section{Appendix}

\end{document}